# Investigation of Softer Lattice Dynamics in Defect Engineered GeTe Crystals

**Saptak Majumder[1], Pintu Singha[1], Sharath Kumar C.[1], Mayanak K. Gupta[2,3], Dharmendra Kumar[4], R. Mittal[2,3], D. K. Shukla[5], M.P Saravanan[5], Deepshikha Jaiswal-Nagar[1] and Vinayak B. Kamble[1,‡]**

[1]School of Physics, Indian Institute of Science Education and Research Thiruvananthapuram, India 695551.

[2]Solid State Physics Division, Bhabha Atomic Research Centre, Trombay, Mumbai 400085, India.

[3]Homi Bhabha National Institute, Anushaktinagar, Mumbai 400094, India.

[4]Department of Physics, Central University of Punjab, Badal - Bathinda Rd, Ghudda, Punjab 151401, India

[5]UGC-DAE Consortium for Scientific Research, Indore 452001, India

**Abstract.** The impact of Ge vacancies on the low-temperature lattice dynamics of single-crystalline GeTe was investigated through a comparative study of two off-stoichiometric samples: $Ge_{0.8}Te$ ($S_1$) and $Ge_{0.88}Te$ ($S_2$). X-ray diffraction confirms their highly oriented crystalline nature mainly along the $h0l$ plane, while temperature dependent Raman spectroscopy reveals pronounced anharmonicity in $S_1$, indicated by stronger three-phonon scattering in the in-plane E-mode. A suppressed Raman feature at 239 $cm^{-1}$ in $S_2$ suggests fewer disordered $GeTe_{4-n}Ge_n$ tetrahedra, correlating with reduced Ge-Ge bonding signatures. Machine-Learned Molecular Dynamics (MLMD) simulations show dominant Te contributions below 100 $cm^{-1}$, while Ge dominates above, particularly influencing the 120 $cm^{-1}$ mode affected by defects at the Ge site. Complementary calculation of phonon linewidth via MLMD and Temperature Dependent Effective Potential (TDEP) methods affirms the predominance of three phonon scattering below 300 K. Specific heat measurements, modeled using Debye-Einstein formalism, show lower Debye temperatures ($\theta_D$) of 172.3 ± 1.5 K in $Ge_{0.8}Te$ and 176.6 ± 1.7 K for $Ge_{0.88}Te$, respectively, confirming defect-induced lattice softening. Electrical resistivity analysis further corroborates this, indicating reduced effective phonon frequencies in $S_1$. Thus, our results establish that higher Ge vacancies lead to softer, and hence more anharmonic lattice dynamics in GeTe, with its relevance in designing superior thermoelectric and phase-change memory applications.

‡ Corresponding author: kbvinayak@iisertvm.ac.in

## 1. Introduction

The basic harmonic model of lattice dynamics has enabled us to understand many of the thermodynamic properties of solids [1]. However, many physical phenomena, including thermal expansion, thermal conductivity, phase transitions, temperature dependence of phonon frequencies and their linewidths, cannot be explained within the limits of the harmonic approximation. These include the phonon-phonon scattering processes that arise from the anharmonicity of the potential energy surface (PES). Three-phonon scattering is usually the dominant scattering process in many materials, is attributed to the third-order anharmonicity of the PES and involves either the decay of one high energy phonon into two lower-energy phonons or the combination of two phonons into a higher energy phonon, while conserving both energy and momentum. Four-phonon processes originate from quartic (fourth-order) anharmonicity of PES. Here, two phonons interact to form two other phonons with different energy and momentum, or a single phonon decays into three other phonons or vice versa, constrained by the conservation of energy and momentum. These scattering processes attributed to PES limit the lattice thermal conductivity. The presence of these scattering processes is usually reflected in the phonon linewidth, and their temperature dependence can be measured using Raman, inelastic neutron, or X-ray scattering techniques [2]. The three-phonon scattering processes result in a linear scaling of Raman linewidth with temperature, while four-phonon scattering contributes an additional quadratic component, causing non-linearity [1–3].

On the other hand, soft lattice dynamics characterized by low-frequency phonon modes and reduced bonding rigidity also play an essential role in determining the thermal, electronic, and vibrational properties of materials. Such dynamics are particularly important in applications that require efficient heat management and tunable electronic behavior, where softer lattices enhance phonon scattering, thus disrupting heat transport in solids [4–6]. This makes materials with a soft lattice structure, such as GeTe-based compounds, very promising for applications such as thermoelectric devices, where maintaining a temperature gradient is critical for converting heat into electricity [7, 8].

Among some well-known thermoelectric materials, GeTe is a group IV-VI chalcogenide with a narrow band gap of about 0.6 eV and a rhombohedral crystal structure with three Ge and three Te atoms in a unit cell near room temperature [7, 9]. The rhombohedral phase of GeTe is a p-type rock-salt crystal with a shear along the [111] [10]. At temperatures around 670 K, GeTe undergoes a structural phase transition from rhombohedral ($\alpha$) to cubic ($\beta$) phase assisted by the soft-phonon mode [10–13]. The bonding mechanism in GeTe is rather unique, which is the reason for its exceptional thermoelectric properties [14–16]. Recently, Wuttig *et al.* have studied the anharmonicity of monochalcogenides such as GeTe, PbTe, and SnTe by pressure-dependent Raman spectroscopy, resulting in a mode-specific Grüneisan parameter given by equation 1

$$\gamma_i = -\frac{V}{\omega_i}\frac{\partial \omega_i}{\partial V} \quad (1)$$

where $\omega_i$ is the vibrational frequency of the $i^{th}$ mode and $V$ is the volume of the crystal. The value of $\gamma$ for GeTe as high as 3 for the transverse optical mode signifies an inherent large anharmonicity in the lattice [17]. This is the primary reason for the low lattice thermal conductivity in GeTe. Further, the bonding in these monochalcogenides was named as metavalent, and the materials were called "Incipient metals". Materials exhibiting metavalent bonding exhibit properties that are vastly different from those of solids with metallic or covalent bonding [18]. A strong lattice anharmonicity is also a signature of metavalent bonding in most group IV-VI and I-V-VI chalcogenides [14–18]. Additionally, the octahedral coordination in GeTe plays a significant role in creating "soft" bonding as compared to tetrahedral coordination. This leads to a low average sound velocity in GeTe [8, 19]. Moreover, the presence of the soft transverse optical mode in GeTe and other chalcogenides results in a lower mean free path of phonons ($\sim 1 - 100$ nm) [11–13, 20–22]. This is also evident from the unusually broad linewidths in the Raman spectrum of highly crystalline GeTe. This signifies a higher phonon scattering rate, which is attributed to its unique bonding mechanism and a large number of inherent Ge vacancies [23].

GeTe has also been widely used in phase-change memory devices and phase change switches because of its reversible phase-change property, that accompanied by a rapid crystallization and a great difference among optical and electrical properties of amorphous vis-á-vis crystalline states [24–27]. Among mid-temperature thermoelectric materials, GeTe is one of the better performers with a high average power factor ($\sim 25$ $\mu$Wcm$^{-1}$ K$^{-2}$) and low lattice part of thermal conductivity ($\sim 2$ Wm$^{-1}$ K$^{-1}$) [7, 28–30]. Compared to other group IV-VI chalcogenides, GeTe has a higher carrier concentration ($\sim 10^{20} - 10^{21}$ $cm^{-3}$), which is due to abundant Ge vacancies that have the lowest formation energy [7, 28–32]. Moreover, due to the highly degenerate nature of GeTe (p $\sim 10^{20} - 10^{21}$ $cm^{-3}$), the total thermal conductivity ($\kappa$) is dominated by the electronic part ($\kappa_e$). Therefore, this limits the achievement of total thermal conductivity suppression [33, 34]. Recently, Li *et al.* have shown that Ge vacancies in GeTe have a significant effect on its band structure, such as band convergence and widening of band gap, thereby enhancing power factor and suppression of bipolar transport [35]. Also, Jiang *et al.* have recently shown that a controlled distribution of Ge vacancies in GeTe by tuning the sintering temperature can lead to multi-dimensional defects, which can scatter phonons over a wide range of frequencies [36]. Yang *et al.* [9] have recently reported Raman active modes arising from defective/amorphous backgrounds in crystalline GeTe, which is unique. They also explored the temperature dependence of Raman modes from 80 K onward to 680 K. However, for the effect of these defective modes to be visible on the overall lattice dynamics, the defect concentration must vary significantly between the samples.

Our study focuses explicitly on this gap and compares both experimental and numerical calculations, including Machine Learned Molecular Dynamics (MLMD) Simulations and Temperature Dependent Effective Potential (TDEP) approach accordingly.

In this study, direct experimental observations of defect-induced changes in the lattice dynamics via temperature-dependent Raman spectroscopy, heat capacity, and resistivity measurements are reported in significantly off-stoichiometric GeTe crystals. Here, the low-temperature Raman spectrum is chosen as a tool for probing the effect of defects on the vibrational properties, which are the fundamental unit of heat transport in solids. However, at higher temperatures, a fine comparison of the Raman spectra would be challenging due to the significant thermal broadening of modes and structural phase transition, along with the peril of surface oxidation. It has been observed from changes in the temperature dependence of the linewidth of the in-plane (basal) $E$ mode that there is higher-order phonon scattering in the crystals, as well as a higher anharmonicity has been quantified through the Klemens model for $Ge_{0.80}Te$. Three-phonon scattering via the phonon linewidth has been confirmed via MLMD simulations and a TDEP-based approach. From the low temperature dynamics of the $A_1^T$ mode, it has been observed that it is sensitive to the vibrations from the disordered $GeTe_{4-n}Ge_n$ tetrahedral units in the more non-stoichiometric analogue, whereas the in-plane $E_1$ mode shows no such anomaly. Additionally, from the specific heat measurements, an attempt has been made to compare the structure rigidity in both crystals by estimating the Debye and Einstein temperatures ($\theta_D$, $\theta_E$). Lowering of the Debye temperature with reduction in Ge stoichiometry has been quantified from specific heat measurements. Furthermore, electrical transport measurements reveal a similar impact on vibrational properties through a reduction in the effective phonon frequency ($\omega_e$) in the more defective and softer lattice.

## 2. Experimental Details

### *2.1. Sample preparation and characterization details*

Lustrous ingots of GeTe were synthesized through melting, followed by homogenization and quenching. Two sets of high-purity Ge and Te (99. 999%), one with a stoichiometric Ge composition and the other with 5% excess Ge, were sealed in vacuum-sealed quartz ampoules (~ $10^{-3}$ mbar). The ampoules were heated to 1233 K for more than 20 hours and then homogenized at this temperature for 10 hours. The melt was then gradually cooled to 923 K over 3 hours and annealed at this temperature for an additional 48 hours. Finally, the ampoules were rapidly quenched in ice water from 923 K.

Crystal structure and phase purity of the samples were determined using a PANalytical Empyrean X-ray diffractometer from a Cu-K$\alpha$ radiation (0.15418 nm) source with a step size of $0.017^o$ in a standard Bragg-Brentano ($\theta - 2\theta$) geometry. Scanning Electron Micrographs (SEM) of the same samples were taken using a Nova NANOSEM 450 Scanning Electron Microscope. The elemental composition of the samples was calculated and quantified from an Energy Dispersive Spectra (EDS) using a 300 KeV FEI TECNAI G2 F30 S-Twin HRTEM. The temperature-dependent Raman spectrum of the samples was performed using a Horiba XploRA Plus Raman microscope in a temperature-controlled stage (make, Linkam, UK, Model THMS600) with a temperature control of ± 0.1 K precision and using a 532 nm excitation source through a long-distance 50x magnification lens and a 2400 gr/mm grating. The sample chamber was purged repeatedly with nitrogen gas before starting the liquid nitrogen flow, and the spectrum was acquired at various temperature set points ranging from 83 to 300 K. The specific heat capacity ($C_p$) of the samples was measured using a PPMS (Physical Property Measurement System) by Quantum Design from 2 K to 300 K using the conventional two tau-method. The temperature-dependent electrical resistivity for both samples was measured from 2 K to 250 K in a Physical Property Measurement System from Cryogenic Limited using a linear 4-probe configuration. A bipolar current of ± 10 mA was sourced via a Keithley 6221 current source, and the voltage was recorded using a Keithley 2182A nanovoltmeter. Further, the difference in the voltages was calculated (to remove the contributions from the thermo-emf), and the final value of resistance was obtained by dividing the resultant voltage by the current. The carrier concentration of the samples was measured using a commercial HEMS measurement system (Nano magnetics HEMS, Israel) at room temperature under a magnetic field of ± 1T. In addition, the Seebeck coefficient of the samples was measured using a custom setup ranging from 150 K to room temperature, whose details can be found elsewhere [37].

*2.2. Computational Framework*

The ab initio molecular dynamics simulations (AIMD) were performed, and subsequent forces and energies have been used to train a neural network for force field generation based on a deep neural network algorithm in the DEEPMD code [38, 39]. Subsequent MLMD simulations were performed with this machine-learned potential using LAMMPS [40]. AIMD simulations were performed from 100 K to 1000 K in intervals of 100 K within the NVT framework on a 2×2×2 supercell. Temperatures were controlled using a NoseHoover thermostat [41] with a time constant of 0.1 ps. An energy cut-off of 600 eV, electronic convergence criteria of $10^{-6}$ eV, and a single k-point at the zone center have been used. The comprehensive AIMD dataset is used to train the neural network force field. A cut-off of 8 Å for neighbouring atom-atom interactions is chosen, and the embedding and fitting network sizes are set to (25, 50, 100) and (240, 240, 240), respectively.

The ab initio molecular dynamics simulations (AIMD) were performed, and subsequent forces and energies have been used to train a neural network for force field generation based on a deep neural network algorithm in the DEEPMD code [38, 39]. Subsequent MLMD simulations were performed with this machine-learned potential using LAMMPS [40]. AIMD simulations were performed from 100 K to 1000 K in intervals of 100 K within the NVT framework on a 2×2×2 supercell. Temperatures were controlled using a Nose-Hoover thermostat [41] with a time constant of 0.1 ps. An energy cut-off of 600 eV, electronic convergence criteria of $10^{-6}$ eV and a single k-point at the zone centre have been used. The comprehensive AIMD dataset is used to train the neural network force field. A cut-off of 8 Å for neighbouring atom-atom interactions is chosen, and the embedding and fitting network sizes are set to (25, 50, 100) and (240, 240, 240), respectively.

The generated force field was benchmarked against AIMD computed forces, energies, pair-distribution function and phonon-density of states (Figure S1 in supplementary information (SI)). An excellent agreement between the machine-learned molecular dynamics (MLMD) simulation and AIMD results has been observed, which validates the machine-learned force field.

To study the temperature-dependent phonon properties, the spectral energy density ($\phi(\vec{q}, E)$) has been calculated using NVT MLMD trajectories on a 10×10×10 supercell of rhombohedral unit cell (5000 atoms). The phonon spectral energy density, $\phi(\vec{q}, E)$ at wavevector, $\vec{q}$ and energy, E is defined as [22]:

$$\phi(\vec{q}, E) = \frac{1}{4\pi\tau_0 N}\sum_{\alpha,k} m_k \left|\sum_{n=1}^{N}\int_0^{\tau_0} \overrightarrow{u_\alpha}\left(\begin{smallmatrix} n \\ k \end{smallmatrix}; t\right).\exp\left[i.\vec{q}.\vec{r}\begin{pmatrix} n \\ k \end{pmatrix} - iEt/\hbar\right]dt\right|^2$$

Where N is the number of unit cells in a supercell ($N = N_1 \times N_2 \times N_3$), summation index $\alpha$ runs over cartesian x, y, and z; index $k$ runs over the number of particles in the unit cell.

$m_k$: mass of $k^{th}$ atom,

$\vec{r}\left(\begin{smallmatrix} n \\ k \end{smallmatrix}\right)$: equilibrium position of the $k^{th}$ atom in the nth unit cell,

$\overrightarrow{u_\alpha}\left(\begin{smallmatrix} n \\ k \end{smallmatrix}; t\right)$ : velocity of the $k^{th}$ atom in the $n^{th}$ unit cell at time $t$.

An MD simulation with a supercell dimension ($N_1 \times N_2 \times N_3$) and trajectory length of $\tau_0$ ps gives an energy and momentum resolution of $\Delta E = 4.136/\tau_0$ meV and $\Delta q = \frac{2\pi}{aN_1}i + \frac{2\pi}{aN_2}j + \frac{2\pi}{N_3}k$, respectively. Here, a is the lattice parameter of the unit cell. We have also computed the temperature dependence by projecting the phonon eigenvectors on MD trajectories followed by a Fourier transform. This allows us to probe the individual phonon mode temperature dependence. While in the SED calculations, it is difficult to track the very close phonon energy temperature dependence.

Three- and four-phonon linewidth calculation: The linewidths contributed from three and four-phonon processes were calculated using the Temperature Dependent Effective Potential (TDEP) approach, which is a perturbation-based calculation [42, 43]. The linewidth is computed using the $2^{nd}$, $3^{rd}$ and $4^{th}$ order renormalized force-constants (FC) at 300 K implemented in TDEP [42, 43]. These renormalized force constants were calculated using AIMD simulations at 300 K. The cut-offs for the $2^{nd}$, $3^{rd}$ and $4^{th}$ force constants were 10, 7 and 5 Å respectively.

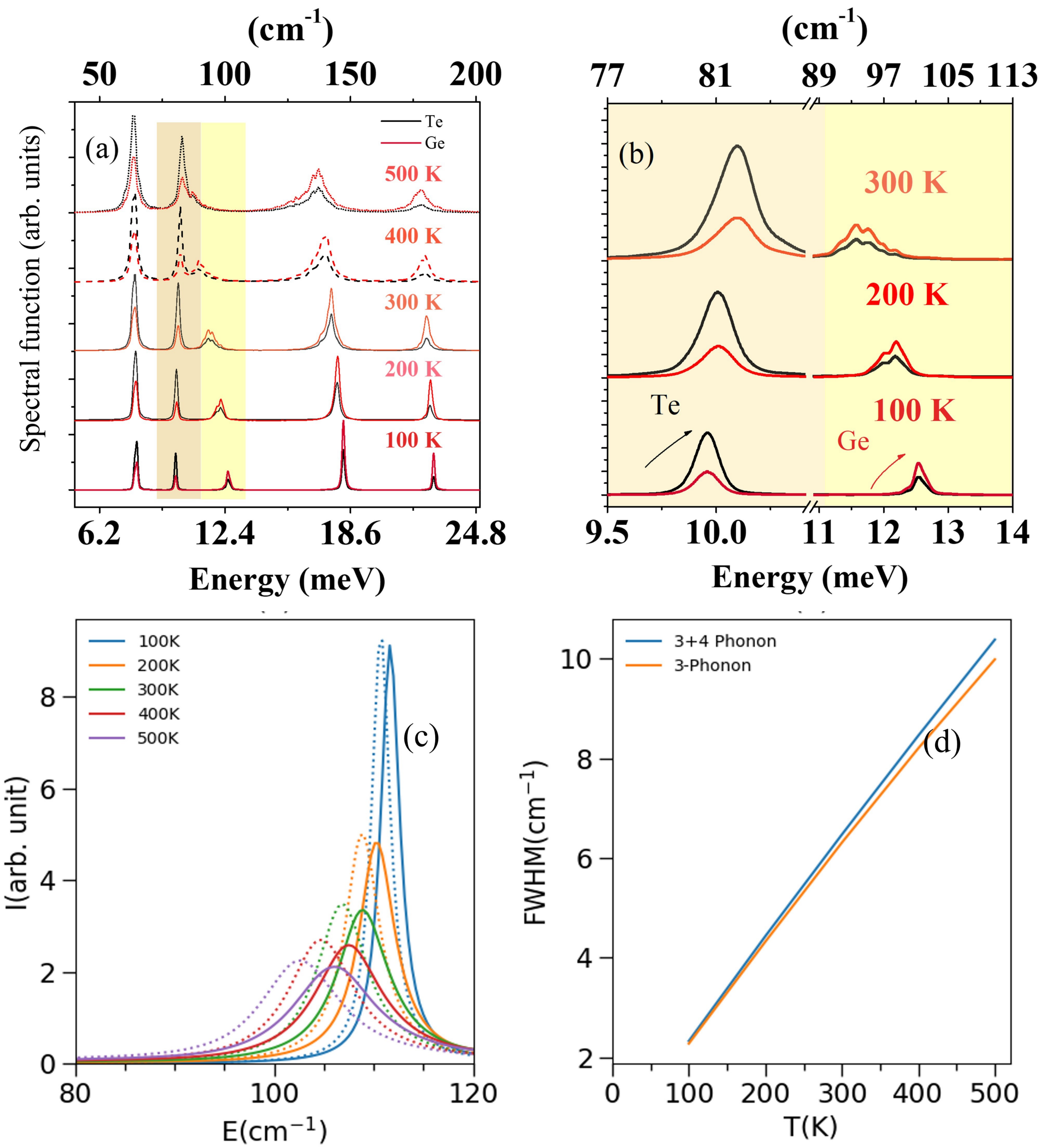

**Figure 1.** (a) The calculated phonon spectral function separately for Ge and Te in GeTe at the zone-centre and their evolution with temperature obtained using MLMD simulations. (b) Magnified portion from 80 $cm^{-1}$ onwards highlighted according to the onset of Ge spectral dominance over Te. (c) TDEP-based calculation showing the variation of E mode for GeTe, considering three phonon (dotted line) and three + four phonon scattering (solid line). (d) Linewidth of E mode (FWHM in $cm^{-1}$) considering three phonon processes only (orange line) and considering both three + four phonon processes together (blue line).

## 3. Results and Discussions

### *3.1. Temperature dependent Phonon Spectral functions using MLMD simulations and TDEP approach*

The calculated phonon spectral function for Ge and Te atoms individually for GeTe at various temperatures from 100 K to 500 K is shown in Figure 1a. The temperature evolution signifies broadening of the modes with temperature, along with softening of certain modes. Also, the heavier Te atoms are observed to dominate the low-energy region and hence the low wave-number region. However, beyond 100 $cm^{-1}$ or ~ 12.4 meV (as 1 $cm^{-1}$ = 0.124 meV), a transition in spectral dominance occurs, where the Ge sublattice contributes more significantly, as shown in Figure 1b. Additionally, using the MLMD simulated MD trajectories, the renormalized phonon energies were calculated by projecting the phonon eigenvectors on trajectories, followed by a Fourier transform, where the softening of phonon energy and broadening in phonon linewidth with temperature have been observed. However, during MLMD calculations, all the orders of lattice anharmonicity are usually taken into account, which highlights the true anharmonicity of the system. Therefore, to validate the extent of four-phonon scattering alongside the three-phonon processes, TDEP calculations are performed for the E mode in GeTe, considering only three-phonon processes and three + four-phonon processes, respectively, as shown in Figure 1c. The three-phonon scattering mostly leads to softening of phonon energies with temperature, while four-phonon scattering processes harden the phonon energies [44, 45]. Hence, there will be a competing effect, and the dominating scattering channel will decide the net softening or hardening of a specific phonon mode. However, both processes will increase the linewidth of the phonon mode. Thus, by comparing the temperature dependence of the linewidth arising from three-phonon processes with that from combined three- and four-phonon scattering, one can identify the presence of higher-order phonon interactions. The calculated data from the TDEP approach further validated that the phonon scattering processes till ~ 300 K are dominantly a three-phonon process. However, a slight deviation in the calculated linewidths between the two different scattering processes occurs 400 K onwards as shown in Fig. 1d. The additional plots showing the variation of spectral function with temperature using MLMD and TDEP approach can be found in the SI (refer to Figures S2, S3 in section 2 and S3 in section 3). Therefore, MLMD simulations show a clear temperature-dependent softening and linewidth broadening of the phonon modes. Furthermore, the phonon linewidths calculated from $2^{nd}$, $3^{rd}$, and $4^{th}$ order force constants using TDEP confirm that four-phonon processes only contribute to the linewidth which may reduce the lattice thermal conductivity, albeit at temperatures beyond room temperature (the details regarding the estimation of the $2^{nd}$, $3^{rd}$, and $4^{th}$ order force constants are discussed in section 3 along with Table S1 of SI).

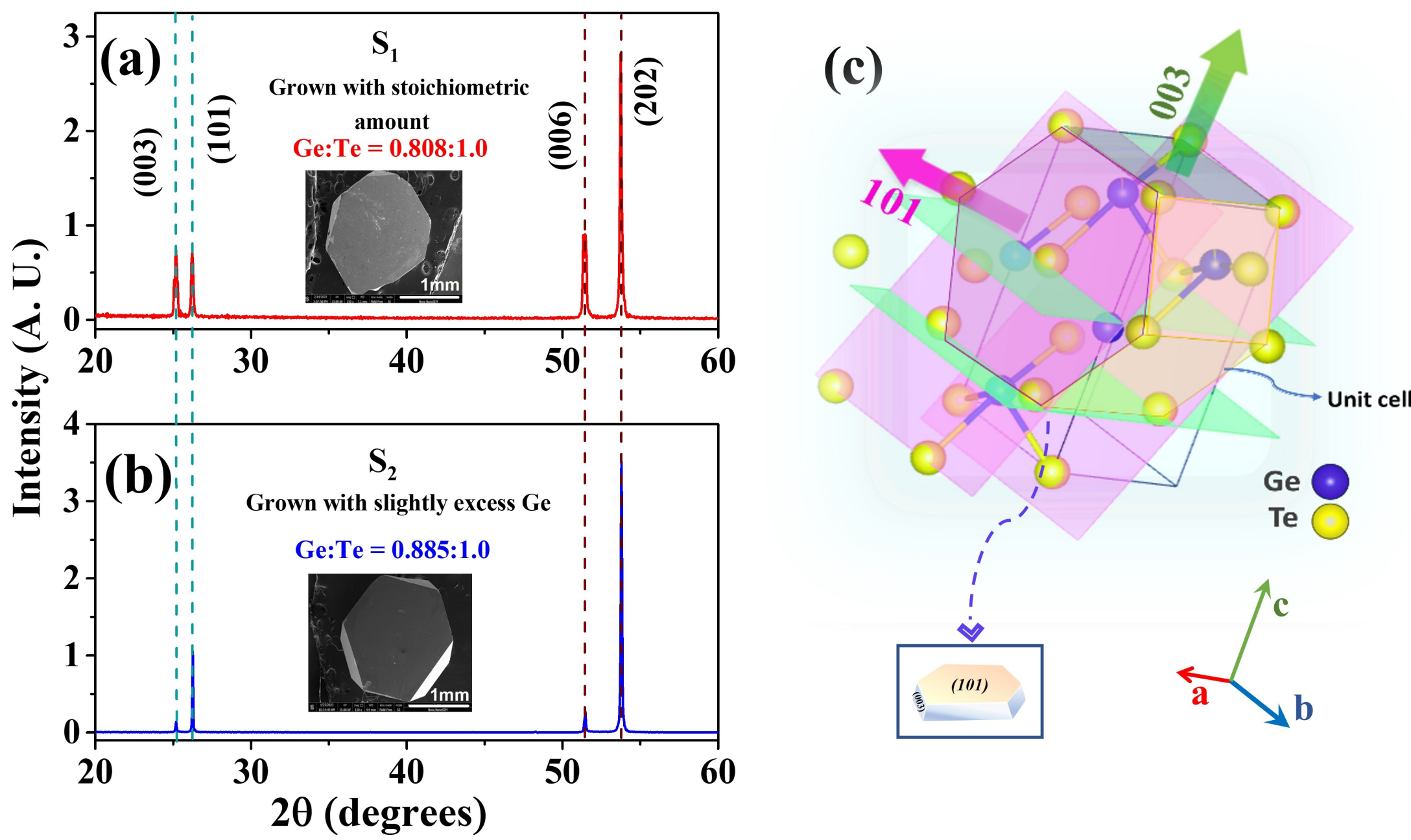

**Figure 2.** XRD patterns of the two faceted GeTe crystals (a) $S_1$ and (b) $S_2$. The SEM images of the crystals with a 1 mm scale bar is shown in the inset, respectively. (c) The schematic of the two crystal planes as seen in the XRD is superimposed with the crystal facets. The inset shows the top plane and prismatic plane indices.

### *3.2. Sample Crystallinity and Crystal Structure*

X-ray diffraction pattern of the shiny ingots (as shown in the SEM images in the insets of Figure 2(a) and 2(b)) depicts the crystalline orientation and phase purity of the samples. The X-ray diffraction patterns of the same crystals show the presence of only two orientations, namely (003) and (101) as shown in Figure 2(a and b). The Bragg positions of the samples were indexed with those of the rhombohedral structure ($R\bar{3}m$, ICDD-04-003-2515) of GeTe. The absence of any other peak in the XRD of the ingots confirms the highly oriented nature of the as-prepared samples. It is also observed from a comparison of the XRD pattern of the two crystals that the intensity ratios of 101:003 and 202:006 are much higher for crystal $S_2$ than for $S_1$. This may be attributed to the different degrees of twinning of the planes in the two crystals, which is an estimate of the extent of twinning in terms of the number of twin planes per unit volume. In our case, the crystal $S_1$, which has more defects, is likely to have more twin planes (to accommodate the strain) and hence shows lesser preferred orientation due to repeated twin boundaries. The (111) family of planes usually acts as twin boundaries in GeTe as they are the natural mirror/symmetry planes of the rhombohedral distortion [46]. In addition, portions of the ingots were finely ground, and their details of the powder XRD pattern signify that the rhombohedral phase is intact even in the powder samples, as shown in Figure S5 in the SI section. The details obtained from the powder XRD refinement of the crushed ingots are added in the SI (Table S2).

A minute impurity phase of Ge is observed in the powder form of crystal $S_1$, which is grown in an exact stoichiometric environment, while the same could not be detected for the crystal $S_2$, although it is grown in a 5% excess Ge environment. Theoretical calculations by Liu *et al.* [47] indicate that Ge vacancies readily form, since they are thermodynamically favored with a lower formation energy in a relatively Te-rich environment. Additionally, Edwards *et al.* have shown that Ge vacancies have a formation energy that is one-third of that of Te vacancies. Furthermore, Ge vacancies do not induce localized gap states, but rather delocalize states just above the valence band, giving rise to degenerate hole-doping-induced metallic conductivity [48]. Thus, crystal $S_1$ is expected to have more Ge vacancy defects as well as higher hole density than crystal $S_2$.

### *3.3. Temperature-dependent Raman Spectroscopy*

To examine the local crystal bonding, environment, and the overall lattice dynamics, temperature-dependent Raman spectroscopy was performed on the freshly cleaved surfaces of the two crystals. The spectra were collected by varying the sample temperature from 83 to 300 K, as shown in Figures 3(a and b). Upon deconvolution of the Raman spectra, a total of five peaks were fitted for $S_1$ and $S_2$, as shown in Figure 3(c) to (f). The fitted peaks are near 90 $cm^{-1}$, 125 $cm^{-1}$, 140 $cm^{-1}$, 159 $cm^{-1}$ and 239 $cm^{-1}$ which corroborate with the observed Raman spectra for rhombohedral GeTe [49, 50].

According to the literature, the peaks near 90 $cm^{-1}$ and 125 $cm^{-1}$ arise due to the vibration of the Ge-Te bond in octahedral coordination, which corroborates the peak positions of the phonon modes at 11.97 meV (96.47 $cm^{-1}$) and 17.98 meV (144.91 $cm^{-1}$) from the MLMD simulations. The energy of the latter is slightly higher than experimental data. The former is due to the doubly-degenerate $E$ mode arising from the vibration of the Ge and Te sublattice along the basal (a-b) plane, whereas the latter is the non-degenerate $A_1$ mode triggered by the vibrations along the three-fold symmetry c-axis [12]. Furthermore, the peak near 140 $cm^{-1}$ is due to the vibration of Te-Te bonds (induced by Ge vacancies), and the mode near 159 $cm^{-1}$ is observed as a less intense peak, being a part of a broad background for the 140 $cm^{-1}$ mode. This mode has been assigned due to the vibration of short-range Te chains arising from the breaking of the translational symmetry. The energy of this mode exactly matches the vibrational energy of amorphous Te. Therefore, it is referred to as the vibrational modes due to disordered Te chains [50–54]. Finally, the broad peak near 239 $cm^{-1}$ is due to the antisymmetric stretching of Ge-Ge bonds in defective $GeTe_{4-n}Ge_n$ tetrahedral units [55] (denoted by * in Figure 3(c, d, e, f)) whose intensity is observed to sustain throughout the entire temperature range in $S_1$ unlike $S_2$ [refer to Figures S6 and S7 from SI]. The latter two vibrational modes mainly arise due to localized deviation from ideal crystallinity. The observed broadening of linewidths and reduced mode intensities further suggest that GeTe, in contrast to typical crystalline materials, may exhibit structural heterogeneity, the manifestations of which can be elucidated through temperature-dependent investigations.

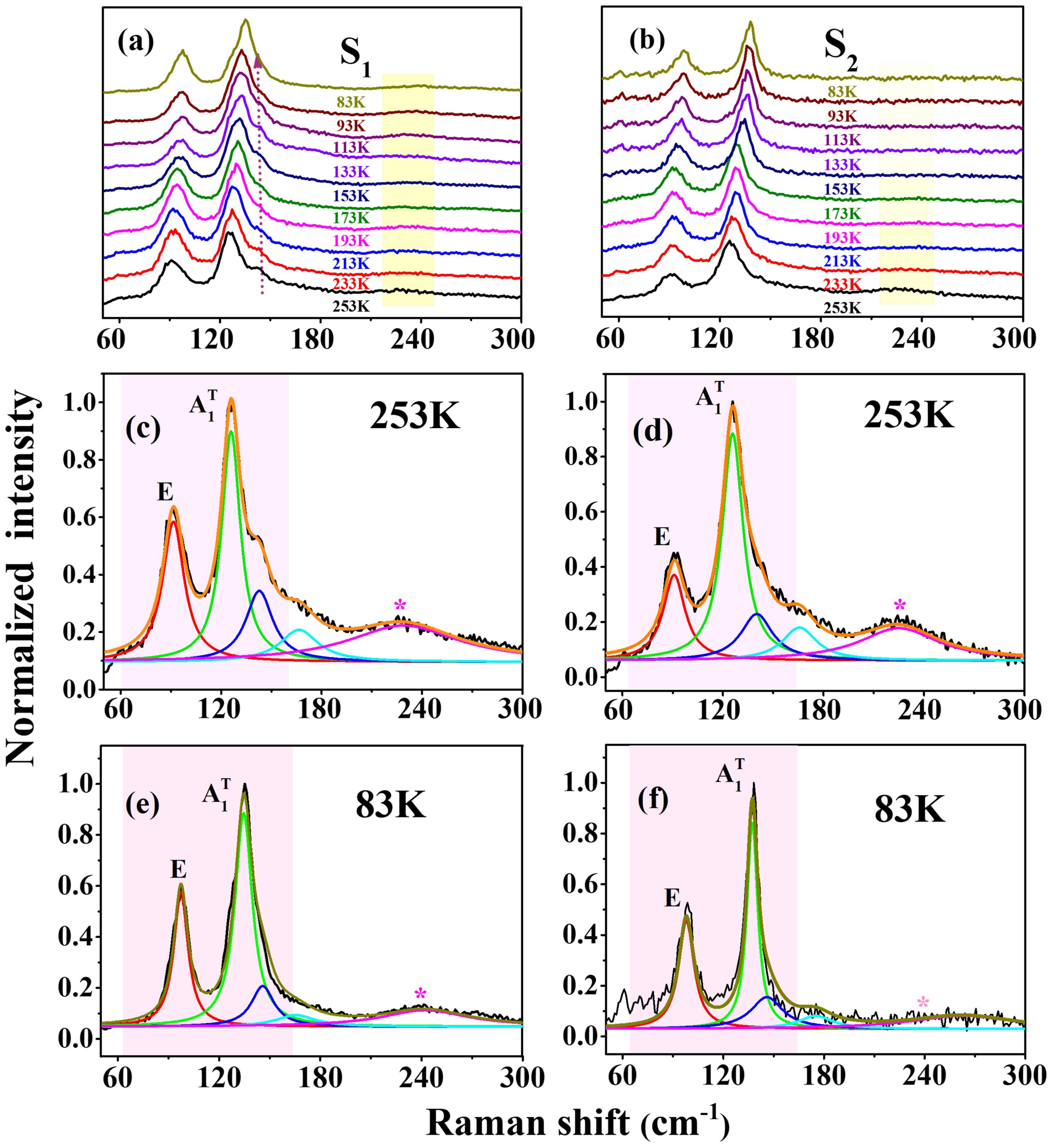

**Figure 3.** The temperature-dependent Raman spectra for crystal (a) $S_1$ and (b) $S_2$ with representative Raman spectra showing different modes deconvoluted for (c) $S_1$, (d) $S_2$ at 253 K, and at 83K for (e) $S_1$ and (f) $S_2$. (* denotes the mode from disordered $GeTe_{4-n}Ge_n$ units)

From the Raman spectra of 253 K, it is evident that sample $S_1$ has a more pronounced mode near 140 $cm^{-1}$, which may be attributed to comparatively higher Te interactions induced from Ge vacancies, as shown in Figures 3(c) and 3(d). The relative intensity of this mode is observed to increase slightly with temperature, whereas its relative intensity is less in the case of $S_2$, which is grown in a 5% excess Ge environment. However, in both samples, the Raman signal near 140 $cm^{-1}$ is consistent yet significantly broad and hence partially overlaps with the $A_1^T$ mode. Therefore, increasing the complexity of the analysis in the linewidth. Therefore, the linewidth of the isolated E mode, unaffected by overlapping modes, has been analysed further. Also, the ratio of I($E$): I($A_1^T$) is higher for $S_1$ as compared to that of $S_2$. This may arise due to the contribution from the set of (003) planes in the E mode, which is relatively higher in $S_1$ than $S_2$ in accordance with the XRD of the two crystals.

### *3.4. Features in the E mode near 90 $cm^{-1}$ and Lattice Anharmonicity*

The doubly degenerate E (TO and LO) mode for GeTe arises due to the vibration of the Ge and Te sublattices perpendicular to the three-fold axis [12]. A clear shift towards a higher wavenumber has been observed in this mode for both the samples with the decrease in temperature, as shown in Figure 4(a), indicating the phonon mode softening with temperature due to thermal expansion. From a linear fit of the variation of the peak position of the E mode from 83 to 300 K, it is observed that the rate of softening of this mode in both $S_1$ and $S_2$ has nearly the same value of $\chi$ ($\chi$ = d$\omega$/dT, where $\omega$ is the Raman shift), irrespective of one having a lesser Ge defect. Therefore, the observed difference in the $\chi$ values being very small (< 0.001 cm$^{-1}$ K$^{-1}$) has provoked the authors to also analyse the linewidth of the E mode. Recently, from Raman spectra on pristine single crystals of GeTe, it has been observed that the linewidths are unusually broad compared to GeSe, SnSe, and other binary chalcogenide single crystals [23, 56–62]. This signifies that the phonon scattering rate is significantly larger for GeTe. This has been attributed to its unique metavalent bonding mechanism, which assists Ge vacancy formation via ease of bond breaking (also refer to Figure 4(b) Therefore, a logical conclusion about lattice anharmonicity from a mere comparison of the linewidth would be difficult and inconclusive. Hence, the temperature evolution of the linewidth has been deconvoluted using a Lorentz function profile and fitted using the Klemens model as proposed by Balkanski et. al. [63]. From each set of deconvoluted spectra, the evolution in the linewidth of the E mode (~ 90 cm$^{-1}$) with temperature was calculated. The linewidth was observed to increase with temperature for both samples without any nonmonotonicities. The evolution in linewidth of Raman active modes with temperature is usually a combined effect arising from multiple higher-order anharmonic interactions between phonon vibrations. Each increases the overall scattering rate as calculated using Fermi's Golden Rule [64]. However, defect-mediated scattering cannot be explicitly justified with a temperature dependence and may require even lower temperatures to be decoupled completely from intrinsic lattice anharmonicity. Specifically focusing on anharmonic phonon-phonon interactions, which depend on the phonon population, the temperature dependence of the linewidth (Γ) from 170 K onward was fitted using a three-phonon Klemens model as shown in Figures 4(c) and (d). The linewidth of Raman modes over a broad temperature range is usually explained by [63]:

$$\Gamma(T) = A + B\left[1 + \frac{2}{(e^x - 1)}\right] + C\left[1 + \frac{3}{(e^y - 1)} + \frac{3}{(e^y - 1)^2}\right] \tag{2}$$

where A is the broadening due to inhomogeneous lattice strain and phonon confinement effects, B and C are the coefficients of the three and four-phonon scattering processes in the linewidth, respectively. Additionally, $x = \hbar\omega_o/(2k_BT)$, $y = \hbar\omega_o/(3k_BT)$ where $k_B$ is Boltzmann's constant. Here, $\hbar\omega_o$ is the vibrational energy of the specific mode in the ground state. The value of this zero-point energy was calculated by plotting the Raman shift of the E mode with temperature and extracting the intercept following a linear fit using the equation.

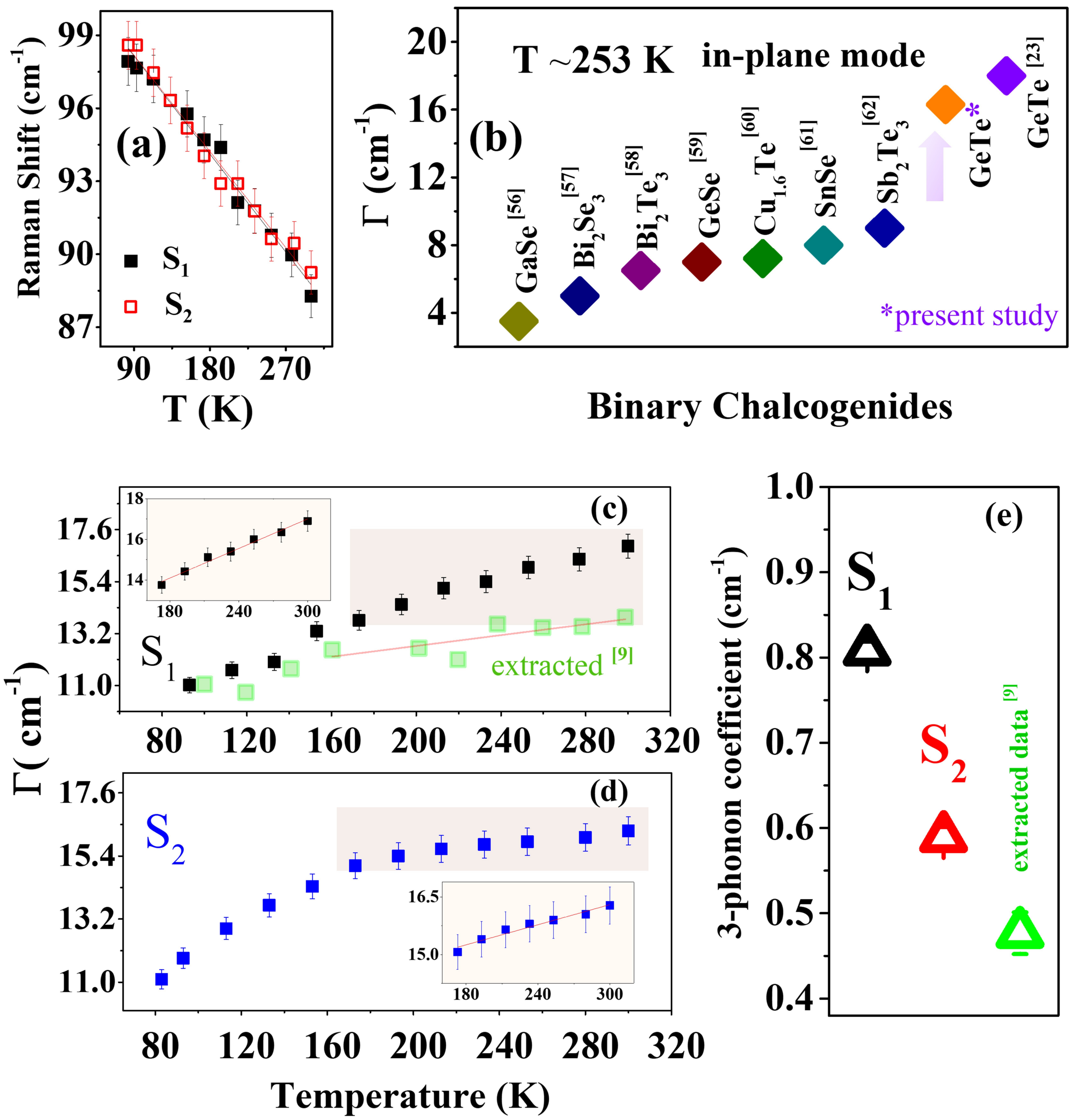

**Figure 4.** (a) Raman shifts vs. temperature plotted for E mode in $S_1$ and $S_2$, along with linear fit (in semi-transparent black and red). (b) Comparison of the in-plane vibration linewidths of similar binary chalcogenides. (c) Linewidths of the E mode for crystal $S_1$, extracted data from reference [9] and (d) crystal $S_2$, along with the fitted data (170 K onwards) using the three-phonon Klemens model in the insets (e) Comparison of the three-phonon coefficient for $S_1$, $S_2$, and extracted data from ref. [9] obtained from the Klemens model, highlighting the change in lattice anharmonicity across samples.

**Table 1.** Parameters obtained using equations (2) and (3) for $S_1$ and $S_2$

| Sample | $A$ (cm$^{-1}$) | $B$ (cm$^{-1}$) | $C$ (cm$^{-1}$) | $\omega_o$ (cm$^{-1}$) | $\chi$ (cm$^{-1}$K$^{-1}$) |
|---|---|---|---|---|---|
| $S_1$ | 10.23(2) | 0.80(4) | 0 | 102.7(3) | -0.04482(4) |
| $S_2$ | 11.74(2) | 0.58(3) | 0 | 101.9(2) | -0.04394(2) |
| ref. [9] | 10.30(7) | 0.47(3) | 0 | 101.3(2) | -0.02955(7) |

$$\omega(T) = \omega_o + \chi T \tag{3}$$

Where, $\chi$ is the first-order temperature coefficient calculated from the slope of $\omega(T)$ versus T plot for a particular Raman mode. The parameters obtained from fitting the linewidth using the three-phonon Klemens model for the two crystals are listed in Table 1 below.

From fitting the linewidth using the Klemens model, it's clear that in the measured temperature range, three-phonon scattering processes contribute significantly. However, from the experimental data, the presence of four-phonon processes is not warranted. Additionally, the linewidth of the E mode from ref. [9] in the similar temperature range has been extracted and fitted using the three-phonon Klemens model, as shown in Figure 4(c) for a better comparison of our experimental data. It is well known that the three-phonon (cubic) term contributes significantly to anharmonicity, while higher-order phonon-phonon interactions can further enhance it. This phenomena is experimentally observed in the linewidth of Raman modes at higher temperature [3, 65–68]. Three phonon processes signify that the optical phonon of energy $\hbar\omega_o$ at the Γ point decays into two acoustic phonons from the same branch while conserving both energy and momentum [63]. The coefficient of the three-phonon scattering processes are shown in Figure 4(e). It has been obtained that the three-phonon coefficient for $S_1$ is sufficiently higher from that of $S_2$ and from the extracted data of ref. [9] in the measured temperature range. However, the obtained value of the three-phonon coefficient is closer for sample $S_2$ and ref. [9]. Therefore, a clear difference in the extent of three-phonon scattering signifies a more anharmonic lattice for $S_1$. Experimentally, a similar effect in the linewidth has been observed in Bi-doped GaAs (LO mode) [69], Cu doped rutile-$TiO_2$ nanorods ($E_g$, $A_g^1$ mode) [70] and most recently in Se-doped $Bi_2Te_3$ in both its $E_g^1$ and $A_{1g}^1$ [71] which has been ascribed mainly due to anharmonic phonon-phonon interactions. However, it has also been shown by Vankayala *et al.* [72] that at higher temperature (> 400 K), four-phonon scattering processes are crucial in lowering the lattice thermal conductivity of pristine GeTe, which is a desired attribute of a good thermoelectric material like GeTe. To theoretically verify this trend in the linewidth, phonon spectral functions using MLMD simulations have been calculated, which show a similar trend in the linewidth as discussed in section 3.1. Additionally, using the TDEP approach, which is are perturbation-based calculation, it is found that for the E mode in GeTe, an additional four-phonon mechanism alongside three-phonon processes begins to slightly affect the phonon linewidth at T∼ 400 K and onwards. Therefore, these calculations further validate our approximation of considering only three-phonon scattering processes in the measured temperature range for modeling our experimental linewidth. GeTe also has a higher phonon band gap (∼10 meV), therefore, fewer decay channels of optical phonon exists and four-phonon scattering probablity [65]. This is in sharp contrast with SnSe, PbSe, etc., where only three-phonon scattering processes dominate thermal transport across a broad temperature region [28, 65, 72–75].

### *3.5. Defect-induced features in the $A_1^T$ mode near 125 cm$^{-1}$*

The nondegenerate $A_1$ Raman active mode in GeTe arises from vibrations of the lattice parallel to the three-fold axis and the $A_1^T$ mode is the transverse component of this vibrational mode [12]. For an ideally stoichiometric GeTe crystal, the structure has only $GeTe_6$ Octahedra as shown in Figure 5 (b). However, these units may be partially transformed to tetrahedra due to non-stoichiometric effects, which mainly affect the Ge sites. Here, the ratio of the number of Ge:Te atoms per tetrahedral unit may vary as $GeTe_{4-n}Ge_n$ units where n=1, 2 and 3[76]. A graphical representation of the GeTe lattice along the c-axis is shown in Figure 5(a), and stacking of the $GeTe_6$ octahedral units with probable Ge defects is shown in Figure 5(b). Figure 5(c) shows the defective $GeTe_{4-n}Ge_n$ (for n = 1, 2) bonding environments as sub-structures of the parent $Te_6$ octahedral cage. Now, from the variation in the Raman shift for the$A_1^T$ mode with temperature in $S_1$ and $S_2$, it has been observed that there is comparatively more stiffening of the $A_1^T$ mode for crystal $S_2$ as compared to $S_1$ at low temperatures. A bifurcation is observed in the low temperature Raman shifts (~160 K onward) between $S_1$ and $S_2$ as a result as shown in Figure 5(d), with a higher differential Raman shift for the $A_1^T$ mode as shown in 5(e). Linear fitting of the Raman shift could not be implemented here for both the samples as the curves have a distinct change in slope. To the best of our knowledge, this anomalous softening/uneven stiffening in the $A_1^T$ mode as shown in 5(d) has been observed for the first time in GeTe. However, it has been previously shown that the variation in the $A_1$ mode with temperature is greater than that of the E mode [12, 77, 78]. Also, the spectral densities as shown in Fig.1b of section 3.1 indicate a higher contribution from Ge as compared to Te for energies above 12 meV (∼100 $cm^{-1}$). Therefore, the Ge defects in our sample are more likely to affect the vibration of the $A_1^T$ mode rather than the E mode.

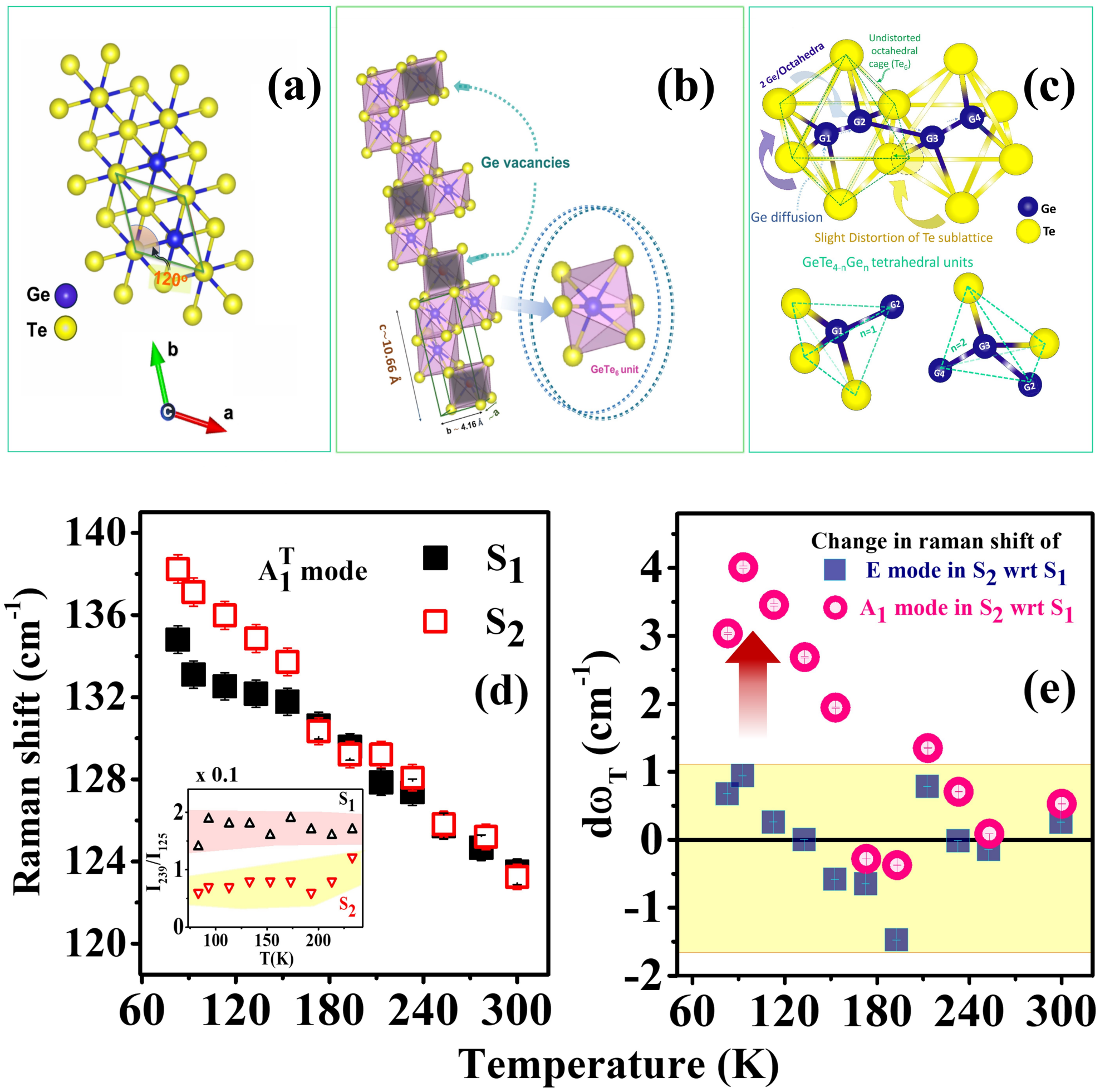

**Figure 5.** (a) Crystal structure of GeTe projected along the three-fold symmetry c-axis. (b) stacking of the $GeTe_6$ octahedra along the c-axis along with an enlarged $GeTe_6$ octahedral unit (c) $Te_6$ octahedral cages with two Ge atoms/ cage along with the possible $GeTe_{4-n}Ge_n$ (n = 1, 2) tetrahedral units (d) Temperature dependent Raman shift for $A_1^T$ mode for $S_1$ and $S_2$ and the ratio of $I_{239}/I_{125}$ for $S_1$ and $S_2$ in the inset (e) Differential Raman shift for $S_2$ with respect to $S_1$ for the E mode and $A_1$ mode.

Although no additional polarizers were used during the experiments, the excitation source—a 532 nm laser—is intrinsically linearly polarized. Furthermore, the Raman spectrometer, particularly the 2400 gr/mm grating and optical collection system, inherently exhibits polarization-dependent sensitivity. Importantly, the twin boundary regions present in the GeTe crystals introduce variations in local crystal orientation, leading to different polarization coupling conditions compared to the normal regions of the crystal. This can alter the relative intensities of Raman modes due to varying Raman tensor components along different crystallographic directions. The observed spectral differences between samples $S_1$ and $S_2$—particularly in the relative intensities of the E and $A_1^T$ modes and the defect-related modes can be susceptible to slight change. However, a systematic polarization-dependent Raman analysis is outside the scope of the current study and may be undertaken under controlled polarization-resolved Raman spectroscopy to quantitatively separate structural from polarization-related contributions to the observed spectra.

### *3.6. Origin of the defective modes in the Raman spectra of GeTe*

Following the crystal structure of GeTe in Figure 5(b), it may be clearly observed that there is a higher density of octahedral units centered by Ge atoms stacked along the c-axis in a single unit cell. Therefore, at lower temperatures, the vibration/stretching of the $GeTe_6$ octahedra becomes increasingly susceptible to interactions with vibrational modes originating from defective bonding environments, due to diminished thermal expansion. For a clearer picture, one can see from the deconvoluted Raman spectra (refer to Figure S6 and S7 in section 5 of SI) that there is a relatively weaker Raman signal around 239 $cm^{-1}$, which is due to the antisymmetric stretching of Ge-Ge homopolar bonding in $GeTe_{4-n}Ge_n$ tetrahedral units [55]. Now, these tetrahedral units are the structural pieces of $Ge_{0.33}Te_{0.67}$ and amorphous GeTe [55, 79]. The crystallization of GeTe occurs via the conversion of the Ge-rich tetrahedral units into Te-rich tetrahedral units. During this process, the coordination of Ge increases which ultimately results in the formation of $GeTe_6$ octahedral units [53]. Wintersteller *et al.* have recently studied the same by XAS measurements and theoretical modeling of amorphous GeTe. Their study reveals that the formation of these tetrahedral units, requiring smaller Ge atoms to diffuse in the $GeTe_6$ octahedral unit, occurs without the formation of other anti-site defects. This makes the process energy efficient. The Ge diffusion occurs via a slight distortion of the tellurium sublattice and a distortion in the already existing central Ge atom site. Therefore, the crystal structure between amorphous and crystalline GeTe is mainly dominated by diffusion of the smaller Ge units in the $Te_6$ octahedral cages [76] as shown in figure 5(c). Also, the intensity and linewidth of the mode ∼ 239 $cm^{-1}$ when compared with both the E and $A_1^T$ mode signifies that the number of these defective units is less. The magnitudes of phonon lifetime for the 239 $cm^{-1}$ mode are nearly an order of magnitude less (refer to section 6 of SI) as compared to the E and $A_1^T$ mode. These features reinforce the fact that the mode arises from defective/disordered bonding environments that lack long-range periodic order. However, it is observed from the dynamics of this mode that there is a consistent Raman signal from crystal $S_1$, whereas it is fully suppressed below 200 K for $S_2$ (see inset of figure 5(d)). This is a major distinction in the Raman spectra between $S_1$ and $S_2$ at lower temperatures (< 170 K), as all the other peaks besides the mode near 239 cm$^{-1}$ could be clearly deconvoluted for $S_2$.

Thus, it may be concluded that the tetrahedral (defective) units have a comparatively higher density of distribution in $S_1$ as compared to $S_2$. These tetrahedral units are not completely decoupled from the octahedral units in the lattice. Rather, these units are part of the parent $Te_6$ octahedral cage as shown in Figure 5(c). Now, crystal $S_2$, being less defective and closer to a stoichiometric composition, exhibits a lower density of disordered units, trending towards pure octahedral coordination. This structural refinement is more clearly reflected in the temperature-dependent Raman spectra. This has resulted in higher differential Raman shifts of $S_2$ with respect to $S_1$ for the $A_1^T$ mode as temperature is lowered below 160 K, shown in Figure 5(d). The Raman shift of a semi-amorphous sample of GeTe before crystallization also shows that the modes mainly arise from the defective-tetrahedral units which is similar to [53] (refer to Figure S8 in SI). Additionally, from Rietveld analysis of the powdered samples of $S_1$ and $S_2$, it has been quantified that there is an order of magnitude change in the c-axis for $S_1$ with respect to $S_2$ ($\Delta c_{(S1,\,S2)}$ = 0.01 Å) as compared to the a/b axis ($\Delta a(b)_{(S1,\,S2)}$ = −0.002 Å). This indicates that the relative percentage change in the c axis is twice that of the a or b axis, thereby reducing the effective strength of interaction along the c axis. Further, theoretical studies on the bonding mechanism and lattice dynamics along and perpendicular to the three-fold symmetry axis for thesdefect-engineereded crystals may give more insights into this interesting phenomenon observed in GeTe or other ferroelectric materials.

### *3.7. Specific Heat Capacity ($C_p$) and Einstein modes in GeTe*

The specific heat capacity of the crystals was measured down to 2 K from room temperature as shown in Figures 6(a) and 6(b), while 6(c) and 6(d) show the experimental and fitted $C_p/T$ versus $T$ using the Debye-Einstein model for $S_1$ and $S_2$ respectively. GeTe being a material with a very high carrier concentration has a considerably high electronic contribution in the specific heat which is evident from the Sommerfeld term for temperatures below 4 K. The Sommerfeld constant ($\gamma$) of the samples was obtained from the intercept of $C_p/T$ versus $T^2$ at temperatures below 4 K and the values are tabulated in Table 2, which are close to the reported values for GeTe [80]. From the plot of $C_p/T^3$ versus $T$ at low temperatures, it was observed that there is a hump-like feature in the data with a maxima near 14 K and the curve around 100 K and onward is flat, following a Debye-like nature at comparatively higher temperatures

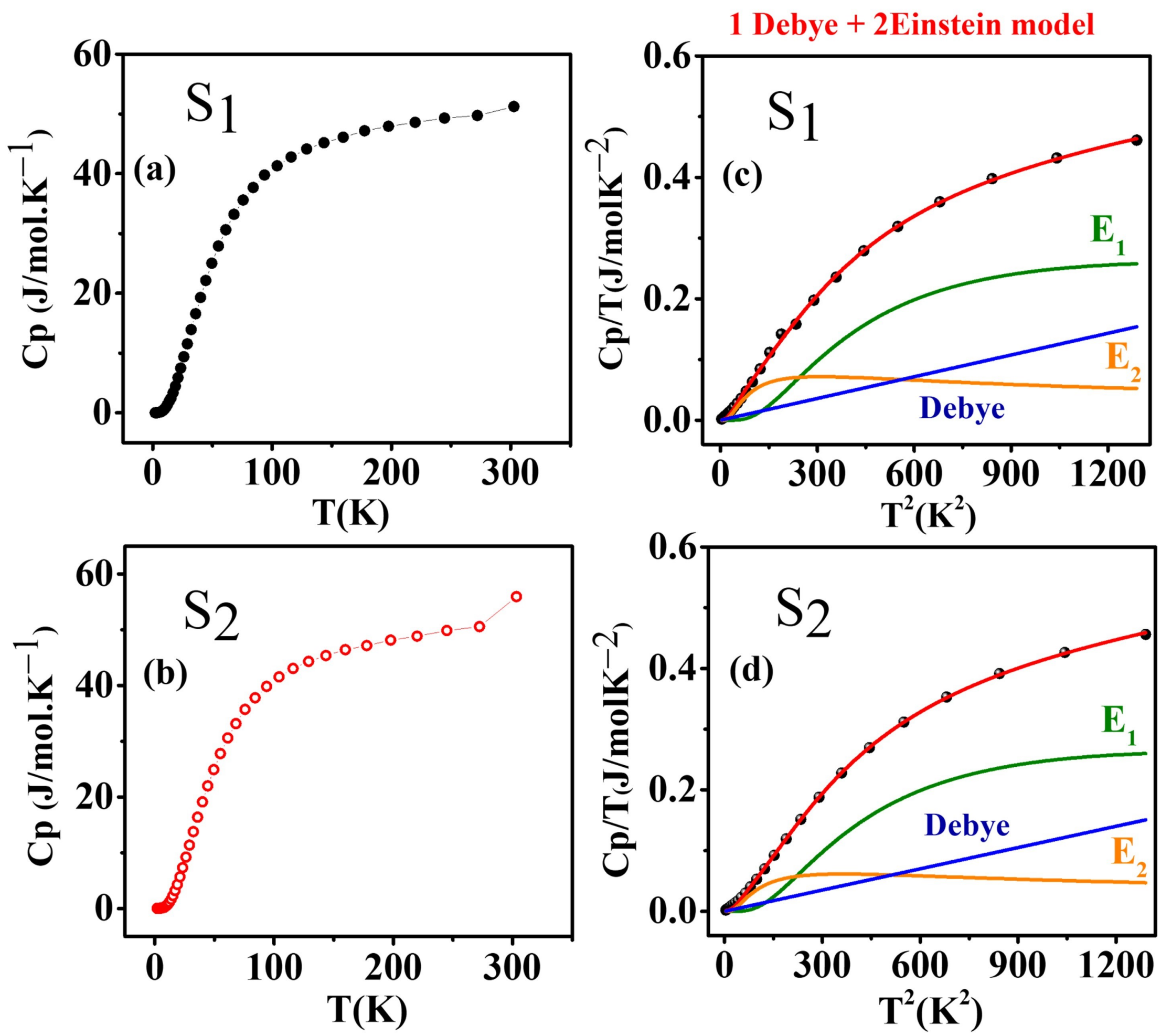

**Figure 6.** (a) Specific heat capacity ($C_p$) as a function of temperature for $S_1$ and for (b) $S_2$, (c) $C_p$/T versus T fitted using the Debye, 2-Einstein model and its deconvoluted components for (c) sample $S_1$ and for (d) for sample $S_2$

(100 K and onward). However, the $C_p$ shows a slight increment beyond the Debye temperature. This increment in specific heat beyond the Dulong–Petit limit arises mainly from strong lattice anharmonicity, low-energy optical phonons that remain active and possible electronic contributions from Ge vacancies. These may ultimately lead to deviations from a usual simple harmonic lattice behavior [81]. Besides, GeTe shows a first-order phase transition at 670 K, where $C_p$ diverges. It is observed from earlier reports that the $C_p$ continues to increase gradually until the phase transition [72]. Similarly, the hump-like feature in the low temperature $C_p/T^3$, as shown in Figure 7, cannot be explained only by Debye's theory, as a pure Debye-like nature corresponds to a $T^3$ dependence in $C_p$, but this observation indicates that some other localized low-frequency vibrational modes might be responsible here [80, 82, 83]. As previously discussed, there are two optical modes (E and $A_1^T$ mode) around 90 $cm^{-1}$ and 125 $cm^{-1}$ in the Raman spectra for GeTe. Therefore, the $C_p$ data were fitted using a 1-Debye and 2-Einstein model as shown in Figures 6(c) and 6(d), which takes into account the two low-frequency optic modes as Einstein oscillators amid a Debye continuum [84]. Hence, the specific heat data was fitted as:

$$C_p/T = \gamma + A_D \frac{12\pi^4 R}{5\theta_D^3} T^2 + 3R\sum_i \left( \left(A_{Ei}(\theta_{Ei})^2 (T^2)^{-3/2}\right) \frac{exp^{(\theta_{Ei}/T)}}{\left[exp^{(\theta_{Ei}/T)} - 1\right]^2} \right) \quad (4)$$

Where the first, second, and third terms correspond to the electronic (Sommerfeld) and the lattice part comprising the Debye and localized Einstein-like oscillator's contribution to specific heat. Here, $\gamma$ is the Sommerfeld constant. R represents the universal gas constant. $\theta_D$ and $\theta_{Ei}$ are the Debye temperature and the $i^{th}$ Einstein temperature, respectively. Also, $A_D$ and $A_{Ei}$ are the pre-factors comprising the product of the number of atoms per formula unit (n) and the oscillator strengths distributed among the Debye and Einstein oscillators, respectively. A slight increment in $A_{E2}$ for sample $S_1$ indicates that the Einstein mode corresponding to $\theta_{E2}$ contributes more to the non-Debye nature of $C_p$ at low temperatures. However, the coefficient $A_{E1}$ is almost the same for both samples.

The fitted data aligns well with the measured $C_p$ values, having $\chi^2 \approx 1.38\times10^{-5}$ and $5.2\times10^{-6}$ for $S_1$ and $S_2$, respectively, indicating that our modeling of specific heat using the combination of 1-Debye and 2-Einstein modes is highly probable for GeTe. Also, the obtained Debye temperature ($\theta_D$) for $S_1$ (172.3 K) and $S_2$ (176.6 K). Recently, it has been shown by several authors that the low-temperature $C_p$ data can be modeled using the Raman spectra/phonon density of states as a reference for identifying the optic modes [82, 83, 85, 86]. Further, the low-frequency optic (Einstein) modes are almost dispersionless and are responsible for increasing the phonon density of states at low energies and henceforth a deviation from the typical Debye-like nature is observed. From a comparison of fitting the specific heat of both the crystals, it is found that crystal $S_1$ has a lower Debye and Einstein temperature as compared to $S_2$, which is more stoichiometric. The Debye temperatures obtained from our experiments for the defect engineered crystals are already on slightly lower side as compared to the reported values for polycrystalline GeTe which is 190 K-200 K. [8] To compare between the two crystals, since there is only slight difference in the Ge:Te stoichiometry i.e. the Ge defect density is the only difference, the subsequent difference in the Debye temperature is going to be subtle. Besides, there is no heavy element substitution involved, so only modest shifts are expected in the characteristic temperature or vibrational energy scales of the lattice (i.e., $\theta_D \sim \omega_D$ and $\theta_E \sim \omega_E$). Materials with a lower Debye temperature usually have soft lattices, meaning their atomic bonds are weaker and more easily distorted [87, 88]. Consequently, these materials tend to exhibit more pronounced anharmonic vibrations, where atomic displacements deviate significantly from a simple harmonic motion [66, 89]. The characteristic frequencies of vibrations, namely $\omega_D$ (Debye) and $\omega_E$ (Einstein) are proportional to the Debye and Einstein temperatures, respectively. Hence, the lowering of $\omega_D$ is a consequence of softer bonding vis-à-vis more anharmonicity due to the distribution of Ge vacancies throughout the GeTe lattice. Lowering of $\omega_E$ indicates that the localized vibrations

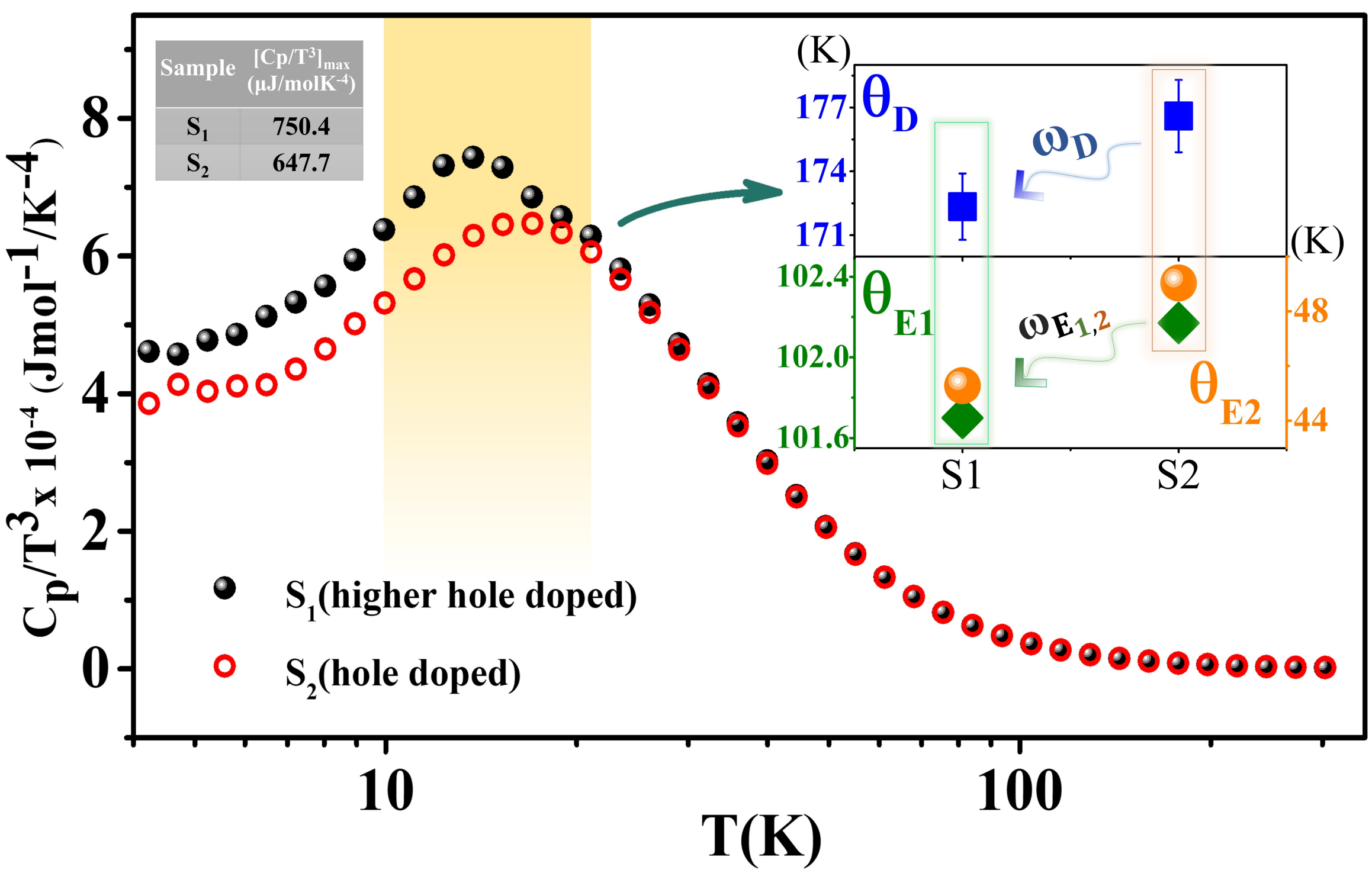

**Figure 7.** Comparison of $C_P/T^3$ vs T for crystals $S_1$ and $S_2$ shows excess contribution for higher hole-doped samples at low temperatures. Also, in the inset (right), lowering of the Debye and Einstein temperatures is highlighted along with frequency softening (represented with decreasing arrows) with (left) the values of the $[C_P/T^3]_{max}$

get easily activated, which also affects the anharmonicity of the lattice by decreasing the restoring forces. Also, from Figure 7, it is observed that the maxima for $C_p/T^3$ are higher for crystal $S_1$ with higher Ge vacancies ($n \sim 1.41 \times 10^{20}\ cm^{-3}$) as compared to $S_2$ ($n \sim 1.15 \times 10^{20}\ cm^{-3}$). Earlier, Shaltaf *et al.* have shown via DFT calculations that a higher hole concentration in GeTe reduces the overall vibrational density of states, which increases the maxima of $C_p/T^3$ versus $T$, signifying a glassier nature [80]. Recently, Moesgaard *et al.* have modeled the $C_p(T)$ of $Ge_{15}Te_{85}$, which is a phase change material having $\alpha$-GeTe and Te as the predominant crystalline phases, using a combination of Debye-Einstein oscillators. Additionally, a continuous substitution of Te by Sb in $Ge_{15}Te_{85}$ has resulted in a monotonic lowering of the maxima of $C_p/T^3$ and a subsequent increment in $\theta_D$ and $\theta_E$ as Sb is slightly lighter than Te [90]. Lowering of the Debye temperature has also been reported recently in Te vacant $Sb_2Te_3$ via $C_p(T)$ and EXAFS [91]. Interestingly, it has also been reported that the sound velocity ($v_s$) and the Debye temperature ($\theta_D$) in GeTe are higher than compared to SnTe and PbTe but the lattice thermal conductivity $\kappa_{\text{lattice}}$ is lowest for GeTe among these mentioned chalcogenides starting from room temperature [8]. Hence, the disorders in the crystal structure and lattice anharmonicity due to the decay of optical modes play a significant role in lattice softening and bringing down the $\kappa_{\text{lattice}}$.

**Table 2.** Parameters obtained from fitting the $C_p(T)$ using Equation 4

| Sample | $\gamma$ $\left(\frac{mJ}{mole.K^{-2}}\right)$ | $A_D$ | $A_{E1}$ | $A_{E2}$ | $\theta_D$ (K) | $\theta_{E1}$ (K) | $\theta_{E2}$ (K) |
|---|---|---|---|---|---|---|---|
| $S_1$ | 0.27(2) | 0.31(3) | 0.69(4) | 0.085(3) | 172 ± 1.56 | 101.7 ± 0.70 | 45.3 ± 0.56 |
| $S_2$ | 0.67(3) | 0.33(3) | 0.70(3) | 0.079(2) | 176 ± 1.7 | 102.1 ± 0.73 | 49.0 ± 0.54 |

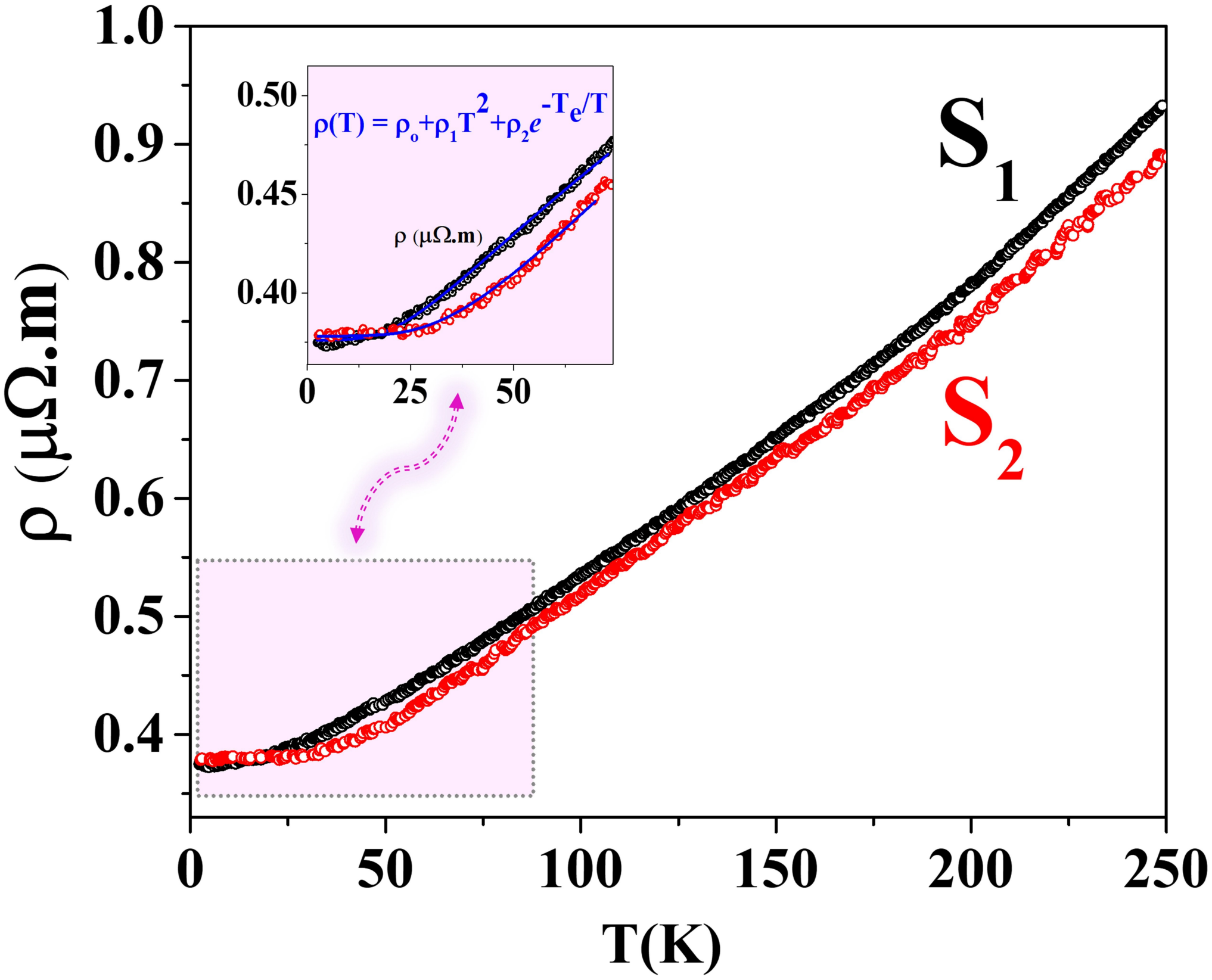

**Figure 8.** Low temperature Resistivity ($\rho$) as a function of temperature (T) for $S_1$ and $S_2$. Magnified low temperature data (highlighted) fitted using equation 5 for $S_1$ and $S_2$ are shown in the inset.

**Table 3.** Obtained parameters from fitting the resistivity of $S_1$ and $S_2$ using equation (5) and the effective phonon frequency from (6)

| Sample | $S_1$ | $S_2$ |
|---|---|---|
| $\rho_0(\Omega.m)$ | 3.76E-7(5) | 3.78E-7(7) |
| $\rho_1(\Omega.m.K^{-2})$ | 4E-12(3) | 1.78E-12(2) |
| $\rho_2(\Omega.m)$ | 2.23E-7(4) | 4.2E-7(5) |
| $T_e$(K) | 81.35(±1.5) | 136(±2.54) |
| $\omega_e$(THz) | 10.62(±1.85) | 17.76(±2.56) |

3.8. Vibrational properties from Electrical Resistivity ρ(T)

The temperature-dependent electrical resistivity measurement for both samples was carried out from 2 to 250 K in a Physical Property Measurement System made by Cryogenic Limited in a linear four-probe configuration. Initially, from the trend in resistivity versus temperature in Figure 8, it is clear that the samples show a metallic behaviour, depicting a degenerate semiconductor. Furthermore, it is observed that the increase in resistivity with temperature for $S_1$ begins noticeably earlier than for $S_2$. This has resulted in a slightly higher value of $\rho(T)$ for $S_1$. To justify this observation, the temperature variation of resistivity $\rho(T)$ was fitted using the equation [92, 93]:

$$\rho(T) = \rho_0 + \rho_1 T^2 + \rho_2 e^{-T_e/T} \quad (5)$$

Where $\rho_0$ is the residual resistivity, $\rho_1$ is the coefficient for the Fermi-liquid term, and the third term arises due to the scattering by low-energy optical phonons or zone boundary acoustic phonon scattering of electrons. Here, $T_e$ is related to the effective phonon frequency ($\omega_e$) via the known relation:

$$k_B T_e = \hbar\omega_e \quad (6)$$

As shown earlier from temperature-dependent Raman spectroscopy in section 3.3 and specific heat capacity measurements in section 3.7, the vibrational properties of GeTe are obstructed by defects such as the Ge vacancies that have the lowest formation energy [47, 48]. After fitting the temperature dependence of $\rho(T)$ at low temperature with equation (5) (see inset of Figure (8)), it has been observed that the value of $T_e$ (~136 K) is higher for crystal $S_2$ than that of $S_1$ (~ 81.3 K). This ultimately results in a higher effective phonon frequency for $S_2$ (~17.76 THz). This signifies a stiffer lattice compared to that of $S_1$ (~10.62 THz), following equation (5). This is concordant with the lowering of the characteristic frequencies from the thermodynamic measurements. Point defects such as vacancies and intrinsic lattice anharmonicity help scatter low-frequency phonons ($\tau_{PD}^{-1} \sim \omega^4$) and hinder heat transport in thermoelectric materials [94]. Therefore, for thermoelectric applications, a defective and yet more anharmonic lattice ($S_1$) signifies that the optical phonons tend to be readily scattered, which may adversely affect the lattice thermal conductivity. Also, as shown in Table 3, the magnitude of $\rho_1$ is rather insignificant compared to that of $\rho_2$. This is due to the fact that the scattering of charge carriers by phonons is the dominating factor over the scattering due to the carriers themselves, affecting the charge transport in the measured temperature range. It has also been shown from transport measurements in single crystals of $Bi_2Se_3$ [95] and $Bi_2GeTe_4$ [92] that the low temperature resistivity $\rho(T)$ follows equation (4) where the exponential term arises mainly due to intervalley scattering from low energy optical phonons or zone boundary acoustic phonons. Apparently, a lowering of the effective phonon frequency has also been quantified via electrical transport measurements on polycrystalline Ni-doped $Sb_2Te_3$ [96], defect-engineered $Bi_2Te_3$ [97] and Te-deficient $Sb_2Te_3$ [91] where the exponential term is also identified to be present in the $\rho(T)$ due to phonon-mediated scattering. Thus, the electrical transport of $S_1$ is not significantly affected due to the abundance of Ge vacancies, although it gives rise to a softer lattice dynamics that could be beneficial for minimizing the thermal conductivity of the solid. Additionally, the effective phonon frequencies estimated from the resistivity measurements reveal a lower value for $S_1$ as compared to $S_2$ [i.e. 10.62(±1.85) THz for $S_1$ and 17.76(±2.56) THz for $S_2$], further substantiating the difference and the trend in Debye temperature.

## 4. Conclusions

To summarize, two GeTe crystals were synthesized that differ substantially in Ge vacancies i.e. Ge:Te stoichiometry. The two samples show a prominent difference in the Raman modes and their temperature dependence. Rhombohedral GeTe mainly has two low-frequency optical modes, around 95 $cm^{-1}$ and 125 $cm^{-1}$, which are dominated with contributions from Te and Ge spectral densities respectively, as shown from our MLMD simulations. However, the existence of other less intense peaks in the Raman spectra signify the complex structural arrangements in GeTe, which is also a well-known phase-change material. Using temperature-dependent Raman spectroscopy, MLMD simulations and TDEP calculations, it is observed that three-phonon processes dominate the phonon scattering mechanism in the measured temperature range, which makes the crystal with more Ge defects to be more anharmonic. However, a gradual difference in the peak positions below 170 K is observed between the two crystals selectively for the $A_1^T$ mode. This additional feature in the Raman spectra has been attributed as due the existence of homopolar Ge-Ge bonding in defective (short range) $GeTe_{4-n}Ge_n$ tetrahedral environments whose fraction is higher for the more defective sample ($S_2$), independent of the temperature range. Following the Raman spectra, the specific heat, $C_P(T)$, is analyzed using a 2-Einstein +1 Debye model to consider the excess vibrational density of states due to the optical modes. This also explains the reason for the non-Debye-like feature observed from the $C_P/T^3$ versus T data. Furthermore, an estimation of the characteristic Debye and Einstein temperature from the $C_P(T)$ data shows an overall lowering of the vibrational energy scales of the lattice for the crystal with a lower Ge:Te ratio ($S_1$). The reduction of the characteristic frequencies of the specific heat capacity is also consistent with the reduction of the effective phonon frequency, which is estimated from low-temperature electrical transport measurements. Overall, via a temperature-dependent study, a comparison of three-phonon scattering has been made, and mode-selective changes in the Raman spectra have been observed. Additionally, specific heat capacity and resistivity data validate the lowering of energy scales of the lattice with the increase in the number of defects for crystalline GeTe. While the differences between the samples are subtle, they are evident and reproducible under controlled synthesis. Future studies may incorporate a larger set of samples with precise defect estimations using techniques such as positron annihilation for statistical verification.

**Acknowledgments**

The authors are thankful to Anusandhan National Research Foundation (erstwhile SERB) for the funding that supported this work. EEQ/2022/001016. The authors also acknowledge the UGC CSR Indore for access to advanced measurement facilities (09/2023/5517 and 07/2023/5134). Additionally, DJ-N acknowledges financial support from SERB, DST, Govt. of India (Grants No. CRG/2021/001262) and IGSTC's WISER program (Grant No. IGSTC/WISER 2024/DJN-2120/45/2024-25/89). Also, the authors would like to thank Dr. Soumya Biswas for the instrumentation involving the low-temperature Raman setup.

All data that support the findings of this study are included within the article (and the supplementary information file).

The supporting information data includes the 1. Benchmarking the machine learned force field using Ab-initio Molecular Dynamics simulations; 2. MLMD-based phonon mode calculations with varying temperature; 3. Temperature-dependent Effective Potential (TDEP) based phonon spectrum; 4. X-ray diffraction of powdered $S_1$ and $S_2$ 5. Deconvoluted Raman Spectra from 83 K to 300 K for $S_1$ and $S_2$; 6. Comparison of phonon lifetimes $\tau_i$ (inverse of Raman FWHM / linewidth) for the 239 $cm^{-1}$ mode with the signature E and $A_1^T$ mode of GeTe crystal; 7. Additional Raman spectra before crystallization; 8. Thermopower Measurements; 9. Low Temperature Specific Heat Capacity analysis; 10. Crystal structure of GeTe showing the arrangement of octahedral $GeTe_6$ units; 11. Raman shifts comparison: Experimental data vs. MLMD simulation; 12. A Dispersion curve for the three-phonon scattering process.

# Supplementary Info

## 1. Benchmarking the machine learned force field using Ab-initio Molecular Dynamics simulations

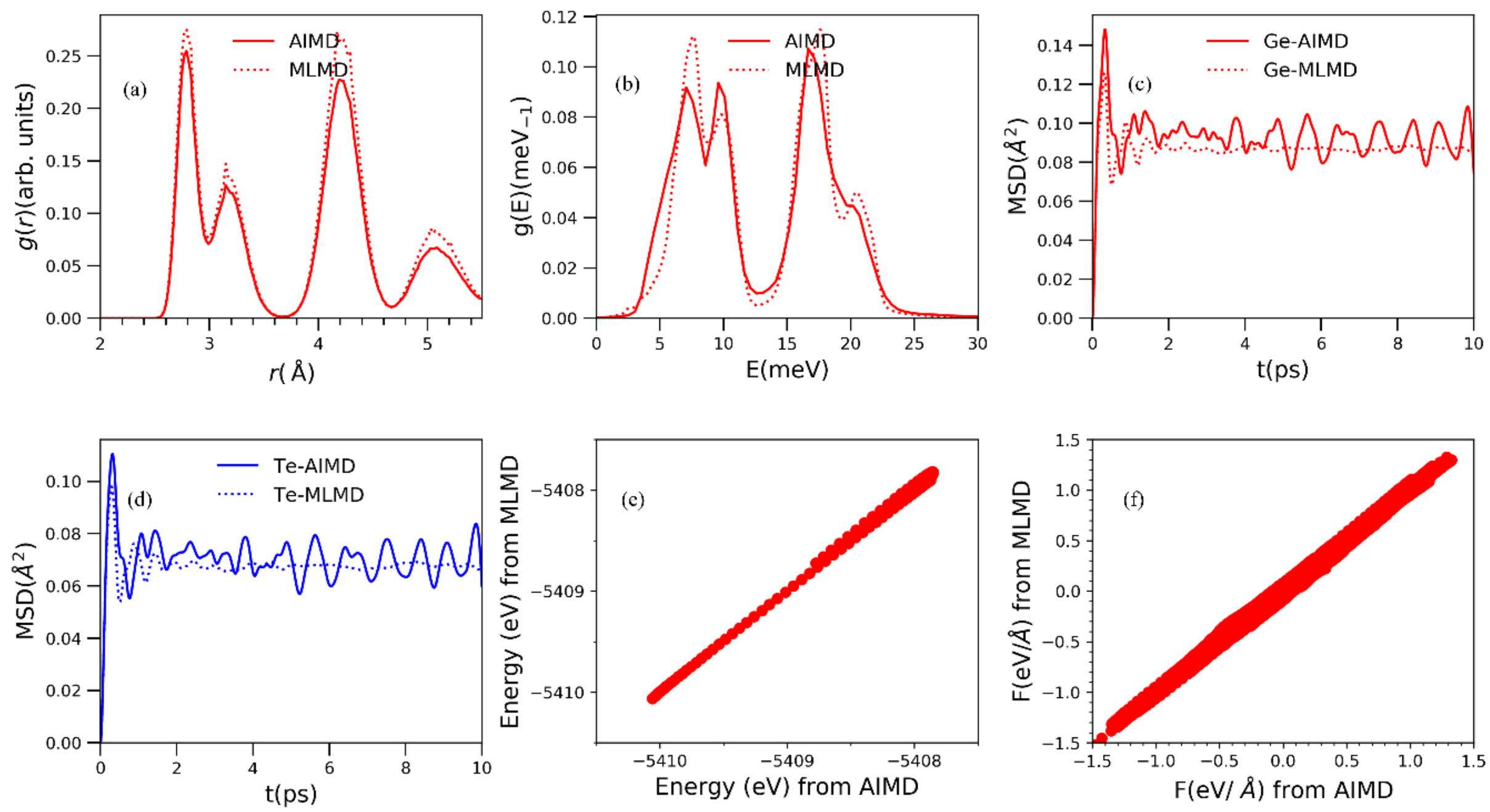

**Figure S1:** The AIMD and MLMD simulations calculated a) pair-distribution function (g(r)), b) phonon-density of states (g(E)), individual mean-squared displacements (MSD) for c) Ge and d) for Te, e) total energies and f) forces. The good agreement between the two simulations validates the machine-learned force field.

Numerical calculations based on Machine-Learned Molecular Dynamics (MLMD) simulations have been carried out to study the phonon spectral functions for GeTe. However, the MLMD simulation requires a machine-learned force field. To validate that, a comprehensive AIMD dataset is used to train the neural network force field. The generated force field has been benchmarked against AIMD computed forces (F), energies, pair-distribution function (g(r)) and phonon-density of states (g(E)). Additionally, the first-principles ab initio molecular dynamics simulations were performed within the plane-wave density functional theory framework, as implemented in the Vienna Ab initio Simulation Package (VASP 5.4.4), where the projector augmented-wave (PAW) method is applied, and the exchange-correlation function takes the generalised gradient approximation (GGA) in the Perdew-Burke-Ernzerhof (PBE) functional [1-3]. A plane-wave kinetic energy cut-off of 520 eV and the self-consistent convergence threshold for electronic minimisation was set to $10^{-8}$ eV for all calculations. The AIMD simulations were performed on a $4 \times 4 \times 2$ supercell of the hexagonal unit cell of R-3m phase (192 atoms). All the calculations were performed within the NVT framework (constant number of particles, volume and temperature, respectively) from 100K to 900K by the steps of 100K for ∼5–10 ps with a time step of 2 fs using a Γ-centered 1×1×1 $k$-point mesh.

An excellent agreement between the machine-learned molecular dynamics (MLMD) simulation and AIMD validates the machine learned force field as shown in Figure S1.

## 2. MLMD based phonon-mode calculations with varying temperature

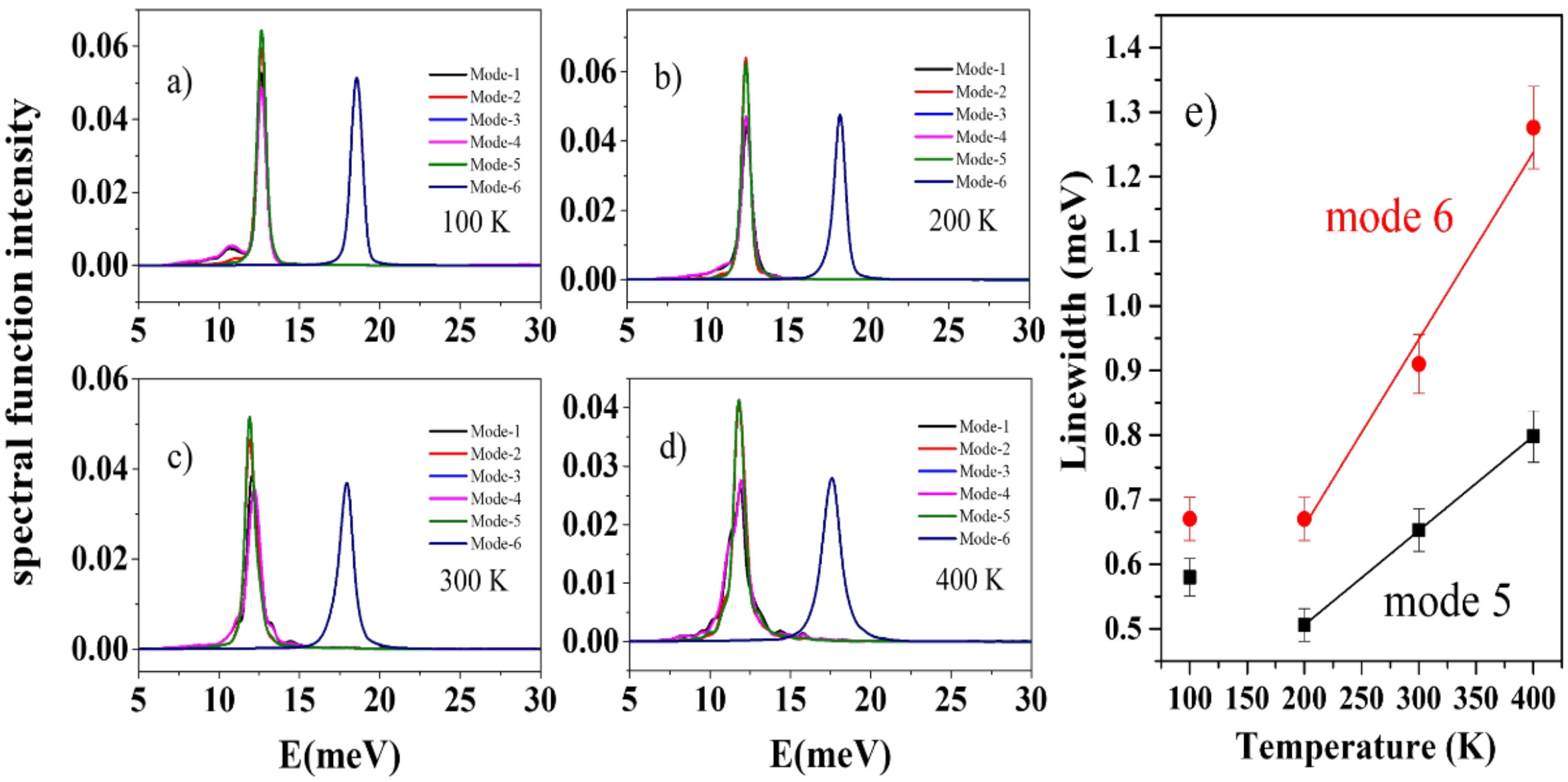

**Figure S2:** Phonon spectral functions calculated using MLMD simulations at a) 100 K, b) 200 K, c) 300 K and d) 400 K. e) Temperature dependance of the linewidths calculated for mode 5 and mode 6 depicting a liner trend (similar to experimental trend).

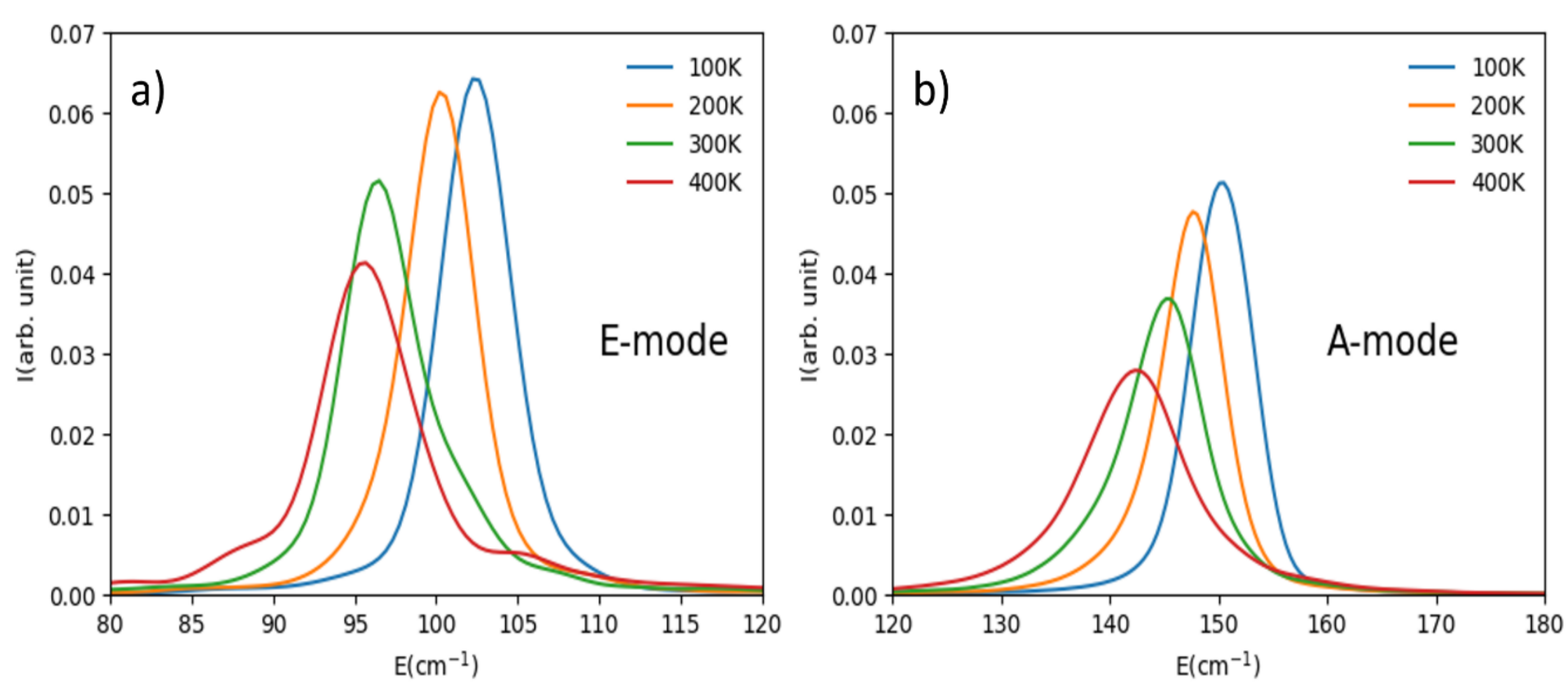

**Figure S3:** Mode-dependent MLMD simulated results showing the softening in peak position of the spectral function with temperature rise.

The overall phonon spectral function demonstrates broadening of the peaks with an increase in temperature as shown in Figure S5 as well as softening of the modes as shown in Figure S6. Using the MLMD based approach, all orders of anharmonicity are taken into account. However, the linewidths calculated for mode 5 and mode 6 (as they nearly overlap in energy with the E and $A_1^T$ mode), both show a linear trend with temperature as shown in Figure S6.

## 3. Temperature-dependent Effective Potential (TDEP) based phonon spectrum

Temperature-dependent effective potential (TDEP) calculations are perturbation based, where one can explicitly demonstrate the difference in phonon linewidths when considering only three-phonon interactions versus the combined effect of three- and four-phonon processes. As shown in Figure S4, the inclusion of four-phonon processes alongside three-phonon interactions results in only a slight increase in the phonon linewidth over the entire vibrational energy spectrum for GeTe (in Thz units), which depicts that additional 4-phonon processes only slightly increases the linewidth and thereby enhance the phonon scattering rate.

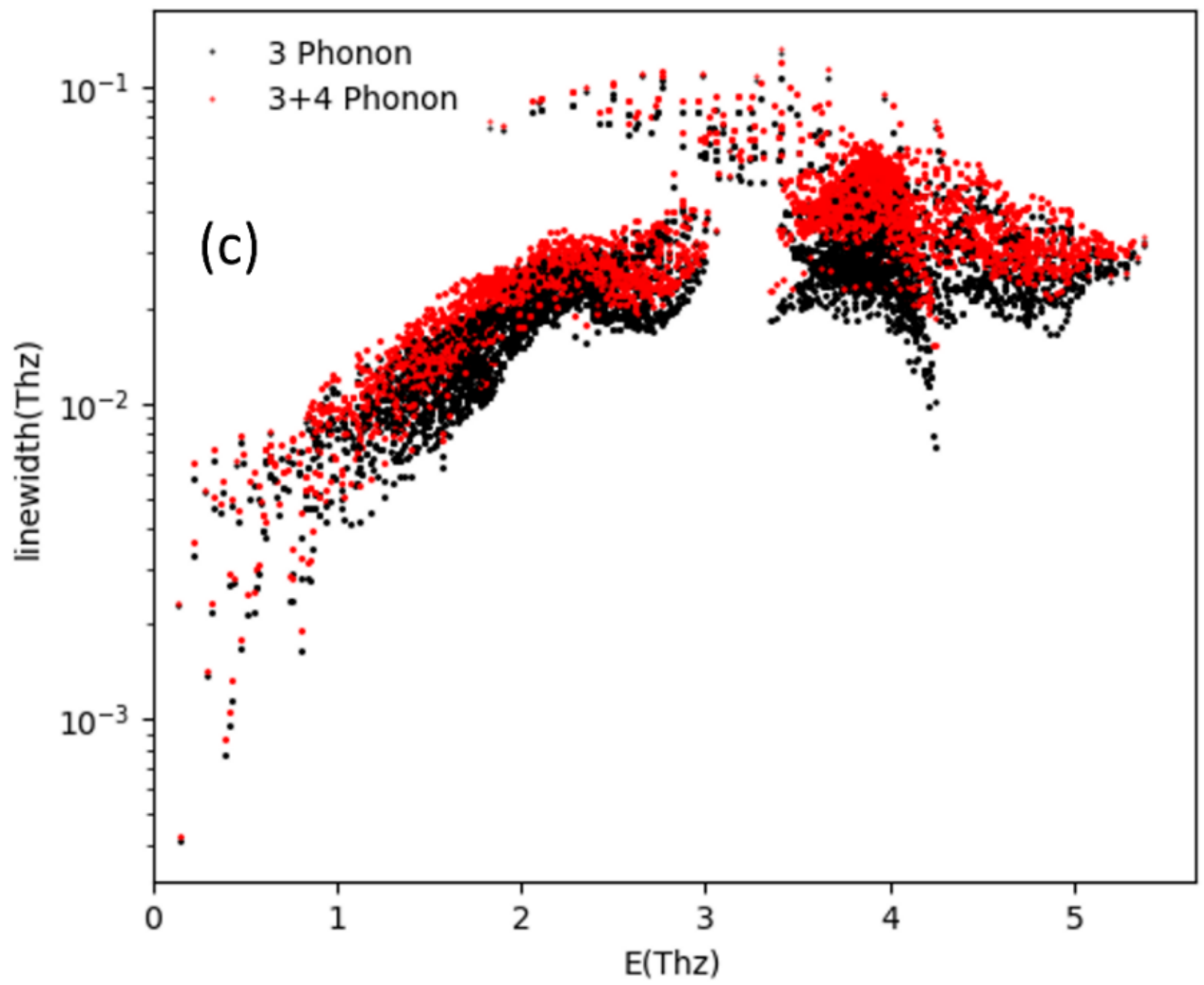

**Figure S4:** The calculated phonon linewidth across the entire vibrational spectrum for GeTe based on from three-phonon and three + four- phonon scattering using TDEP.

This aligns with our experimental observations from the Raman-active E mode, whose linewidth evolution within the measured temperature range is predominantly governed by three-phonon scattering. However, a small fraction of four phonon scattering channels may also exist which is shown by the TDEP calculations. The higher-order force constants were extracted using 5000 snapshots from ab initio molecular dynamics (AIMD) trajectories, employing the TDEP software. To ensure the reliability of the extracted force constants, the cutoff radii (RC) for the second-, third-, and fourth-order terms, were systematically examined as summarized in Table I. The AIMD simulations were conducted with a tight energy cutoff of 600 eV and a stringent convergence criterion of ($10^{-7}$) eV, using the SCAN functional to ensure high accuracy.

**Table SI.** Coefficient of determination (($R^2$)) as a function of cutoff radius for second-order (RC2), third-order (RC3), and fourth-order (RC4) force constants, based on 5000 AIMD snapshots of energy and force data from a 128-atom supercell.

| Cut-off Distance (Å) | 2nd Order (RC2) | 3rd Order (RC3, with RC2 = 8 Å) | 4th Order (RC4, with RC2 = 8 Å and RC3 = 5 Å) |
|---|---|---|---|
| 3 | 0.57 | 0.92 | 0.96 |
| 4 | 0.70 | 0.93 | 0.97 |
| 5 | 0.80 | 0.95 | 0.97 |
| 6 | 0.85 | 0.95 | 0.97 |
| 7 | 0.86 | – | – |
| 8 | 0.86 | – | – |

Based on these results, cutoff distances of 8 Å for second-order, 5 Å for third-order, and 4 Å for fourth-order force constants are found to be sufficient for accurate modeling.

## 4. X-ray diffraction of powdered $S_1$ and $S_2$

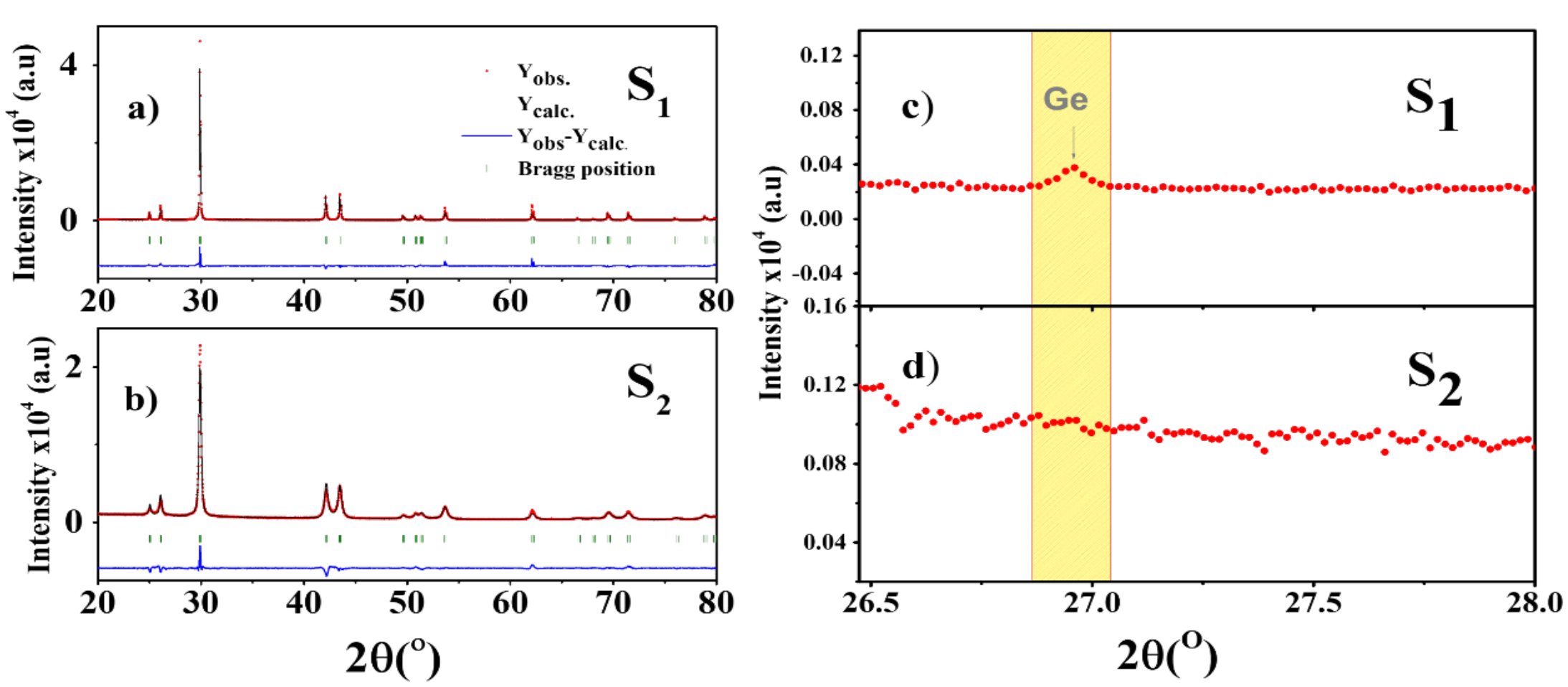

**Figure S5:** Refined Powder XRD pattern for a) sample $S_1$ and b) sample $S_2$ c) Magnified region showing the Ge impurity phase for $S_1$ and d) for $S_2$.

**Table S2.** List of Refined parameters for Crystals $S_1$ and $S_2$.

| **Sample** | **$S_1$** | **$S_2$** |
|---|---|---|
| a or b (Å) | 4.160 (±0.0018) | 4.162 (±0.003) |
| c ( Å) | 10.675(±0.0032) | 10.665 (±0.0036) |
| Cell volume ($Å^3$) | 160.06 (±0.13) | 159.99 (±0.10) |
| $\chi^2$ | 8.0880 | 2.3861 |

The XRD of powdered samples of $S_1$ and $S_2$ is shown in Fig S5 a and b. Figure S5 c and d shows the magnified region from 26.5 $^o$ to 28 $^o$ where a minute peak near 27 $^o$ corresponding to Ge impurity (111 plane) is observed only for the case of $S_1$. The calculated lattice parameter a(b) and c have differences which are of the order of $10^{-3}$ Å and $10^{-2}$ Å respectively. Additionally, for a comparison of our obtained lattice parameters, the changes observed are of the similar order as in ref.[4] where the lattice parameters changes from 4.1624 Å (a) and 10.6762 Å (c) in $Ge_{1.01}Te$ to 4.1696 Å (a) and 10.5816 Å (b) in $Ge_{1.01-x-y}Ti_xBi_yTe$ (x = 0.002, y = 0.008). Similarly, the changes in ref.[5] are also observed to be in the similar order of magnitudes.

## 5. Deconvoluted Raman Spectra from 83 K to 300 K for $S_1$ and $S_2$

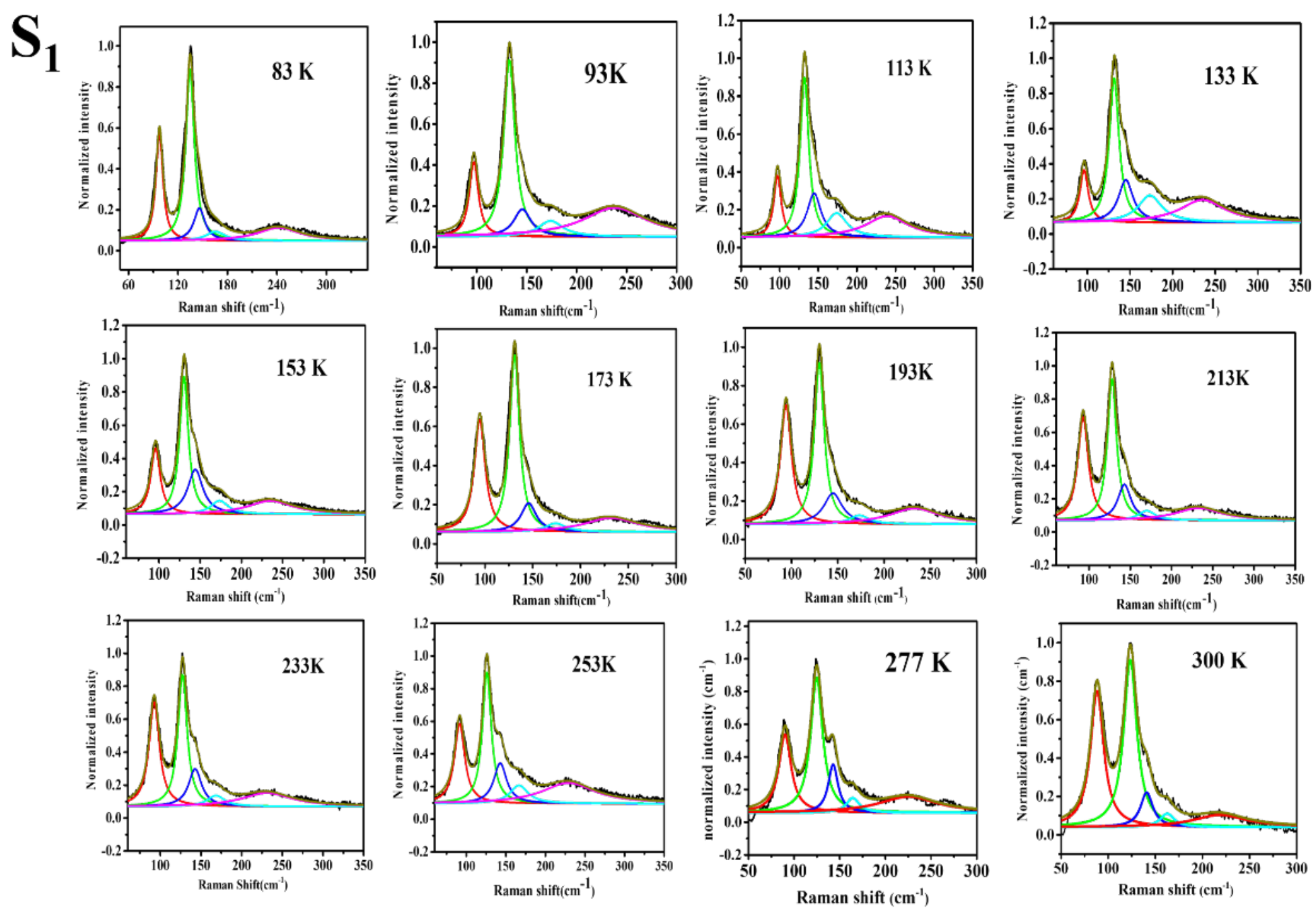

**Figure S6:** The deconvoluted Raman Spectra for crystal $S_1$ collected over the entire temperature range.

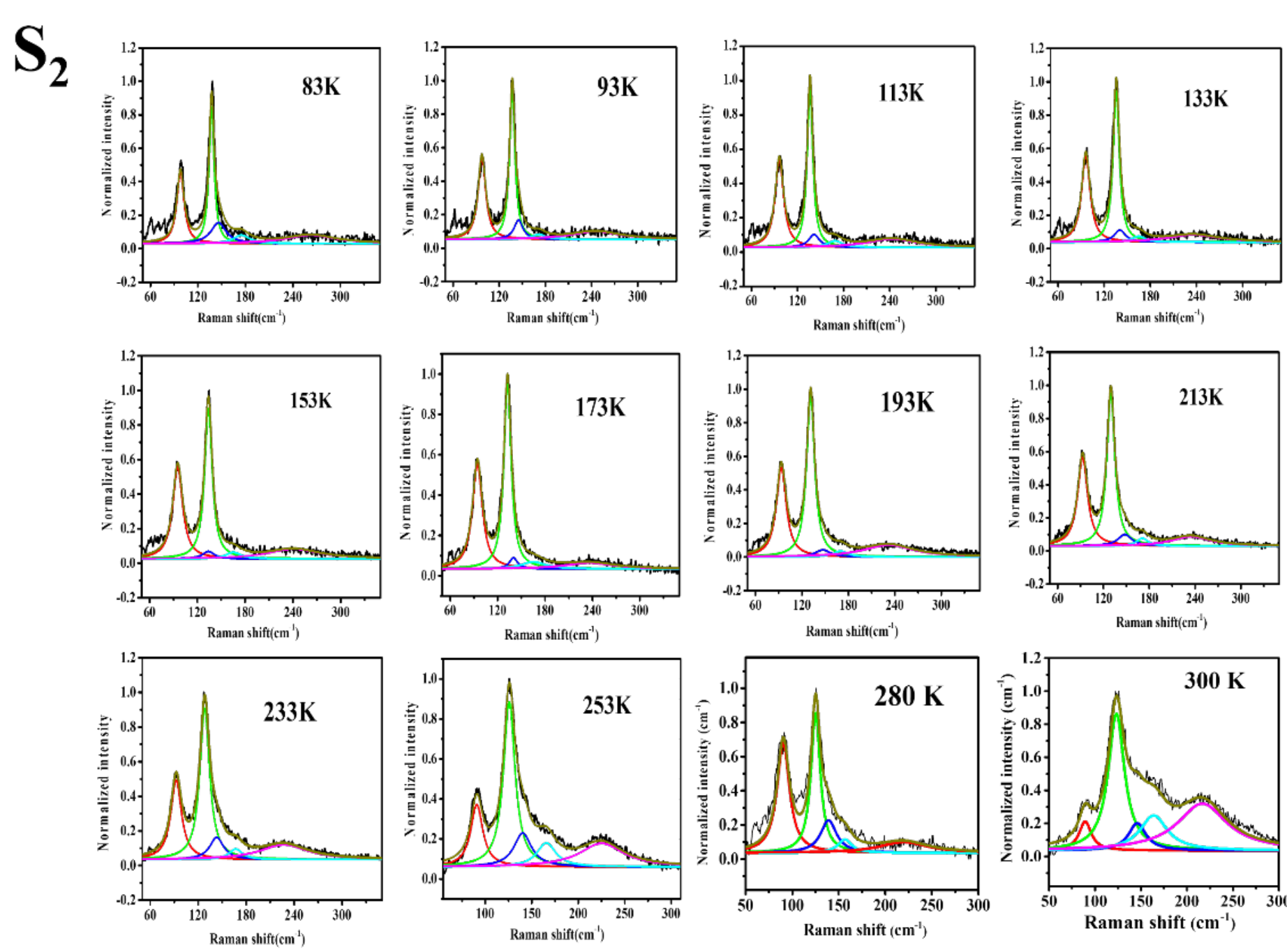

**Figure S7:** The deconvoluted Raman Spectra for crystal $S_2$ collected over the entire temperature range

### 6. Comparison of phonon lifetimes $\tau_i$ (inverse of Raman FWHM / linewidth) for the 239 cm$^{-1}$ mode with the signature E and $A_1^T$ mode of GeTe crystal

The phonon lifetime ($\tau_i$) for the $i^{th}$ Raman-active mode is an estimate of how fast the phonons decay. Therefore, a comparison of the lifetimes between the defective mode and the signature Raman modes for GeTe crystal would verify the less-propagative nature of lattice vibrations coming from short-range defective bonding environments.

As the $\tau_i$ (phonon lifetime of $i^{th}$ mode) is inversely proportional to $\Gamma_i$ (linewidth of the $i^{th}$ Raman mode). Therefore,

At 253 K

**For $S_1$:**

$$\frac{\tau_{239}}{\tau_E} = \left[\frac{\Gamma_{239}}{\Gamma_E}\right]^{-1} = \frac{13.54}{81} = 0.16$$

$$\frac{\tau_{239}}{\tau_A} = \left[\frac{\Gamma_{239}}{\Gamma_A}\right]^{-1} = \frac{15.67}{81} = 0.19$$

**For $S_2$:**

$$\frac{\tau_{239}}{\tau_E} = \left[\frac{\Gamma_{239}}{\Gamma_E}\right]^{-1} = \frac{15}{57} = 0.25$$

$$\frac{\tau_{239}}{\tau_A} = \left[\frac{\Gamma_{239}}{\Gamma_A}\right]^{-1} = \frac{15.8}{57} = 0.27$$

Hence, the 239 cm$^{-1}$ mode having a sufficiently lower lifetime decay faster both for $S_1$ and $S_2$ as compared to the E and $A_1^T$ modes. Also, it signifies that the modes arise from defective bonding environments.

## 7. Additional Raman spectra prior crystallization

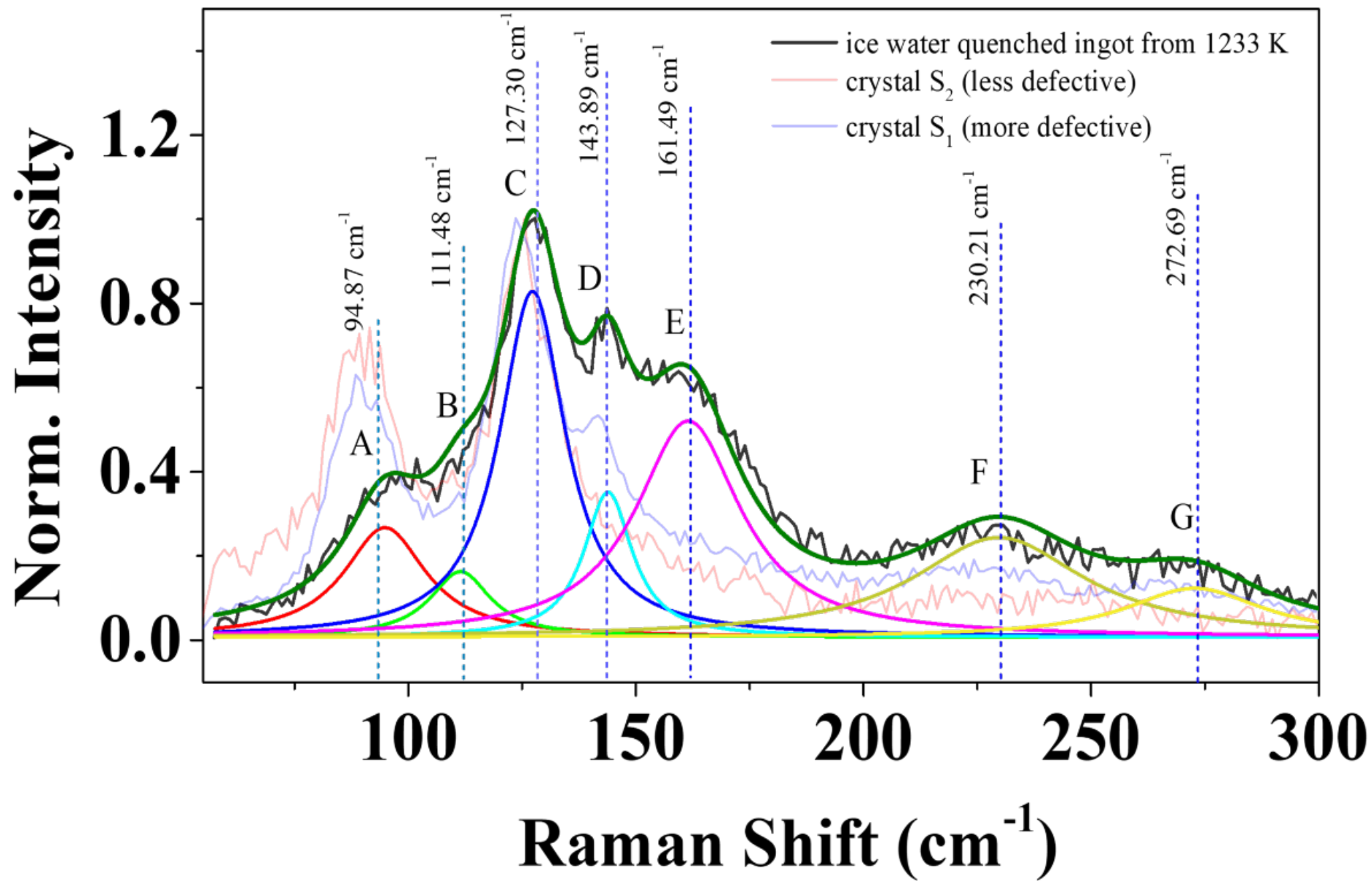

**Figure S8:** Deconvoluted Raman spectra at ambient conditions (in black) of an ice-water quenched ingot before being crystallized. The Raman spectra for crystal $S_1$ and $S_2$ are semi-transparent in blue and red respectively.

During synthesis of these crystals, the crystallization of GeTe occurs when the ampoule is kept for annealing (at 923 K) after homogenizing the melt at 1233 K. During the same process, an ampoule was directly quenched from 1233 K (before annealing) and the Raman spectra of that water quenched sample is showed in figure S6. The water quenched sample being semi-crystalline, shows Raman modes which are substantially broader than the peaks for $S_1$ and $S_2$. Similarly observed by Bastiani et. al [6], the peaks designated as A, B, C, D, E, F and G signify the following as listed in Table S2.

Amorphous GeTe mainly comprises of tetrahedral $GeTe_{4-n}Ge_n$ units where n = 0, 1, 2 and 3. However, crystallization occurs via the rearrangement of these Ge rich tetrahedra into Te rich ones. This process gradually saturates the Ge coordination ultimately forming octahedral $GeTe_6$ [6]. However, minute proportions of these $GeTe_{4-n}Ge_n$ (n = 1, 2 and 3) tetrahedral units may remain coupled inside the $Te_6$ octahedral cages even after annealing the crystals due to non-stoichiometric effects and Ge diffusion [7]. Therefore, the signatures of these defective environments have been observed as peaks of very less intensity in the Raman spectra for the sample $S_2$ and mainly for $S_1$, even though the samples are highly crystalline, oriented and visibly lustrous. For an ideally stoichiometric GeTe crystal, the structure has only $GeTe_6$ Octahedra. However, these units may be partially transformed to tetrahedras due to non-stoichiometric effects mainly arising from substantial vacancies at Ge sites. Here, the ratio of the number of Ge:Te atom per tetrahedral unit may vary as $GeTe_{4-n}Ge_n$ units where n = 1, 2 and 3.

As studied by Bastiani et. al [6], the distributions of these tetrahedral units for amorphous GeTe consist of:

| | |
|---|---|
| ~ 1% | $GeTe_4$ (n=0) |
| ~ 20% | $GeTe_3Ge$ (n=1) |
| ~ 36.75% | $GeTe_2Ge_2$ (n=2) |
| ~ 30% | $GeTeGe_3$ (n=3) |
| ~ 9.15% | $GeGe_4$ (n=4) |

Therefore, considering the distribution of the tetrahedral units which are already sparsely distributed, we can omit n = 0 and 4.

**Table S3:** Peak positions after deconvolution and their attributions based on literature reports of the Raman spectra for amorphous GeTe.

| Peak | A | B | C | D | E | F | G |
|---|---|---|---|---|---|---|---|
| Position | 94.87 | 111.48 | 127.30 | 143.89 | 161.49 | 230.21 | 272.69 |
| Attribution | Bending modes of tetrahedral $GeTe_{4-n}Ge_n$ units + disordered Te chains | Vibrations of corner-sharing tetrahedral units of $GeTe_{4-n}Ge_n$ for n = 0,1 | Vibrations of corner-sharing tetrahedral units of $GeTe_{4-n}Ge_n$ for n = 2,3 | Crystalline Te-Te interactions | Amorphous Te chains | Antisymmetric stretching of the Ge-Ge bond in $GeTe_{4-n}Ge_n$ | Stretching of the Ge-Ge bonds in Amorphous Ge |
| Linewidth ($cm^{-1}$) | 22.1(5) | 16.3(3) | 17.5(2) | 12.9(3) | 28.2 (8) | 46.3(5) | 37.1 (7) |
| Ref. | [8,9] | [10] | [10] | [11] | [9] | [8] | [8] |

**8. Thermopower Measurements**

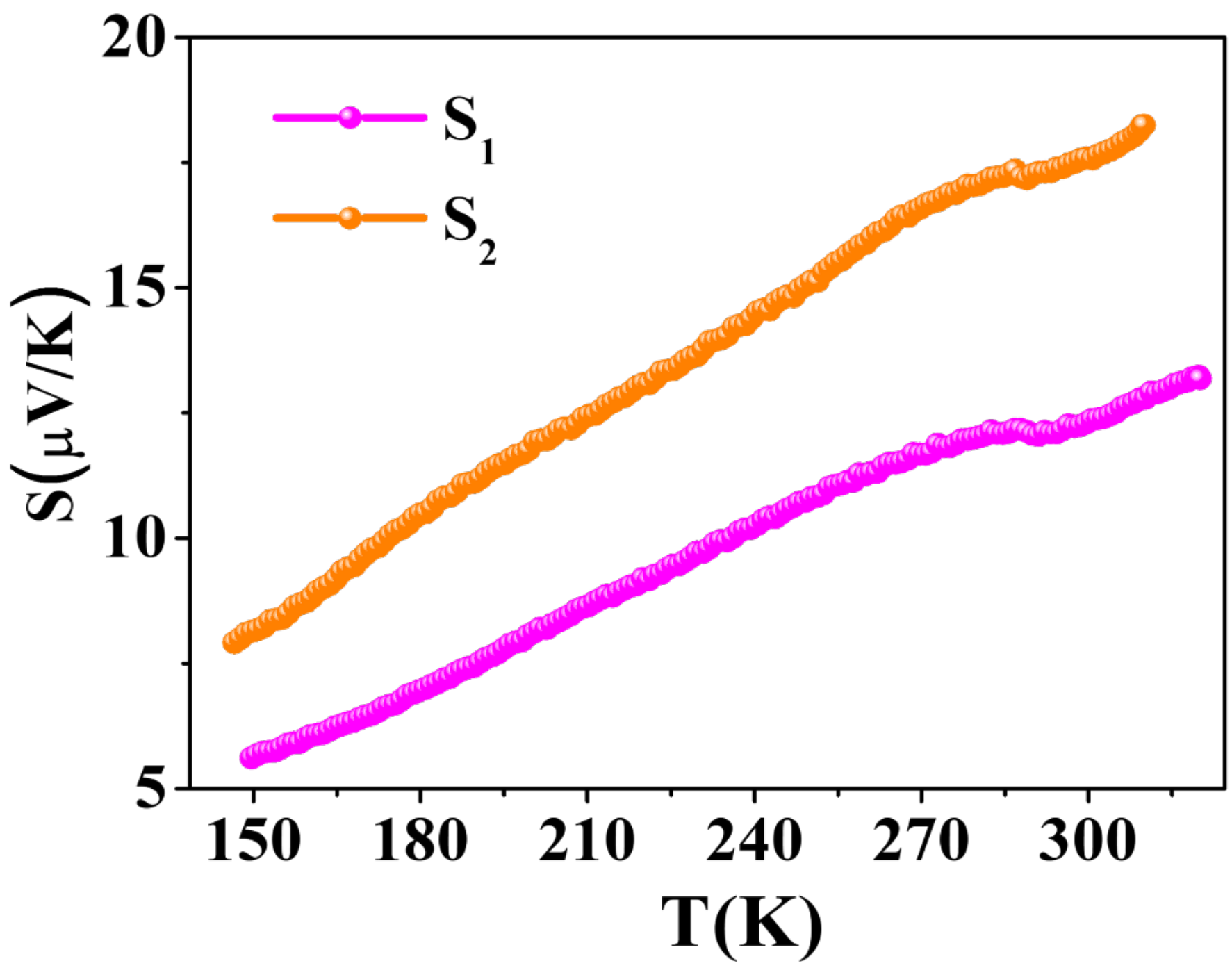

**Figure S9:** The Temperature dependent thermopower data for for crystals $S_1$ and $S_2$.

The thermopower of the samples was measured from 150 K to 330 K and it's clear that the samples are of p-type with holes as majority carriers as shown in Figure S9. Also, the increasing trend in the thermopower depicts the degenerate-type nature of the samples according to the Mott formula. From a comparison of the thermopower of both samples, it is clear that the crystal $S_1$ with more Ge vacancies have a lower value of thermopower as compared to $S_2$. This is mainly due to the reason that a higher concentration of Ge vacancies in GeTe makes it more p-type. Also, each Ge vacancy in the GeTe lattice contributes two holes which increases the overall carrier concentration (n) and according to the Mott formula for *S* [12], the thermopower decreases.

$$S = \frac{8\pi^2 {K_B}^2 T}{3eh^2} m_{DOS}^* \left(\frac{\pi}{3n}\right)^{-2/3}$$

where $K_B$ is the Boltzmann's constant, $h$ is the Planck's constant, $m_{DOS}^*$ is the density of states effective mass and n is the carrier concentration

## 9. Low Temperature Specific Heat Capacity analysis

$$C_P = \gamma T + A_D \frac{12\pi^4 R}{5\theta_D^3} T^3 + 3R \sum_i \left( A_i (\theta_{Ei})^2 (T^2) \frac{exp^{\left(\frac{\theta_{Ei}}{T}\right)}}{(exp^{\left(\frac{\theta_{Ei}}{T}\right)} - 1)^2} \right) \ldots S1$$

Eq. S1 shows the Expression for the Specific heat capacity modelled using the 1-Debye +2-Einstein model

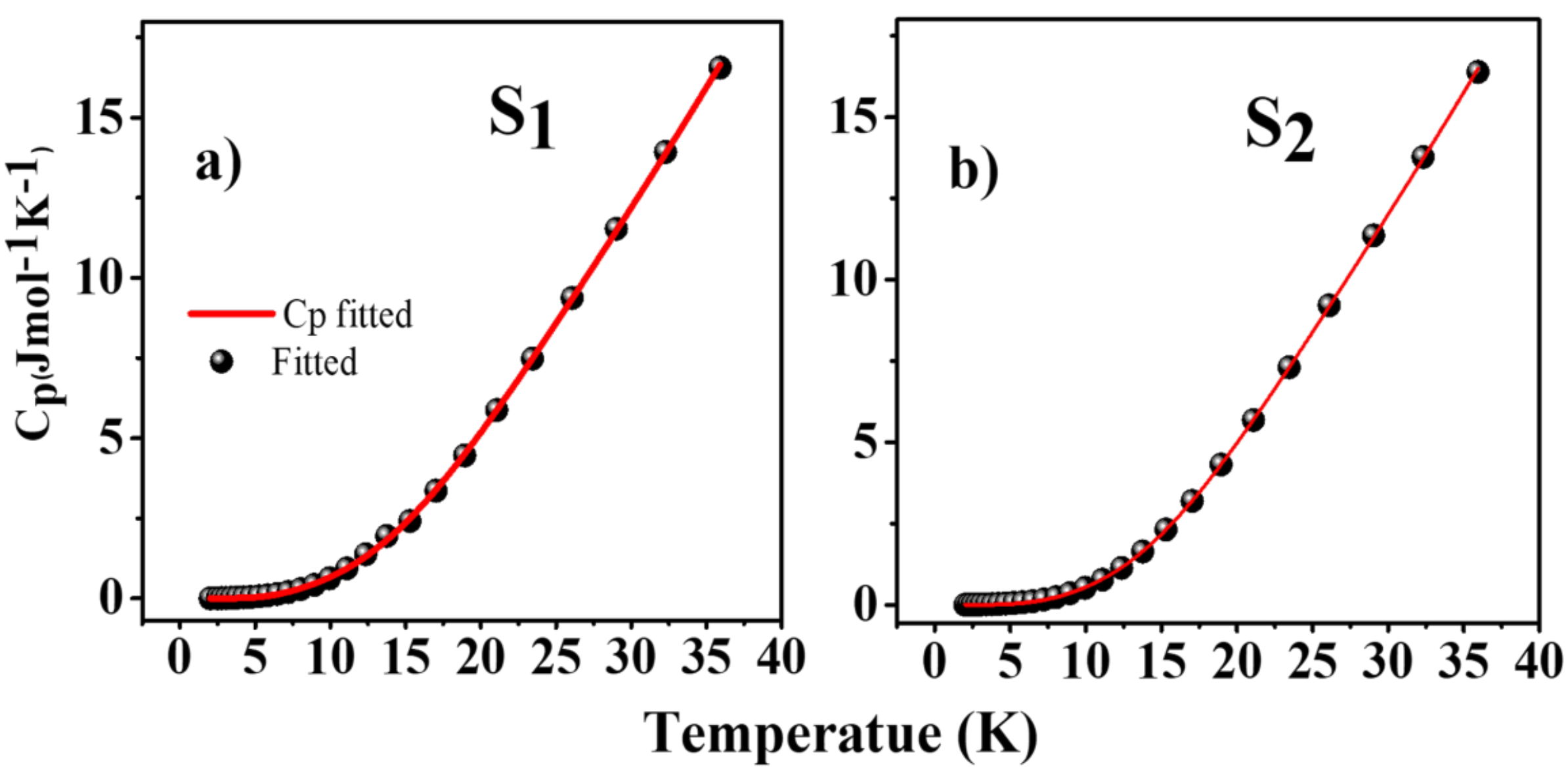

**Figure S10:** The low temperature specific heat, $C_P(T)$ measured and fitted data using equation S1 for (a) $S_1$ and (b) for $S_2$.

## 10. Crystal structure of GeTe showing the arrangement of octahedral $GeTe_6$ units

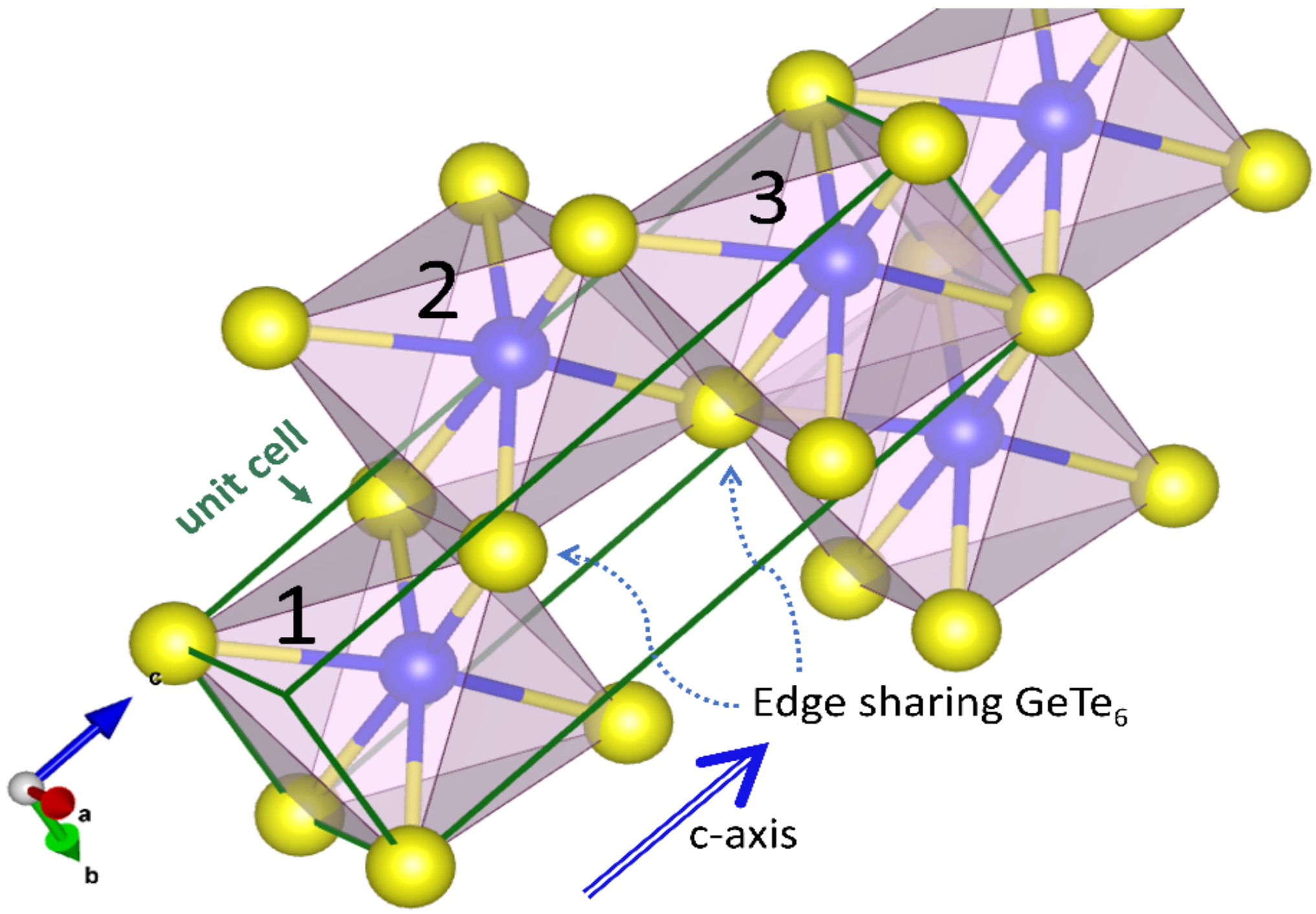

**Figure S11:** Crystal structure of GeTe (Ge in blue and Te in yellow) considering pure octahedral co-ordination (along with the unit cell marked with green boundary)

The crystallographic representation in Figure S11 is illustrated for a pristine GeTe (Ge in blue and Te in yellow) considering pure octahedral co-ordination. Here the octahedral units ($GeTe_6$) are stacked as edge-sharing units (1-2 and 2-3) along the c-axis. Therefore, the number of $GeTe_6$ octahedral unit/ unit cell in the crystal is more along the c-axis.

## 11. Raman shifts comparison: Experimental data vs. MLMD simulation

The phonon modes have been calculated using the MLMD simulations at 100 K, 200 K and

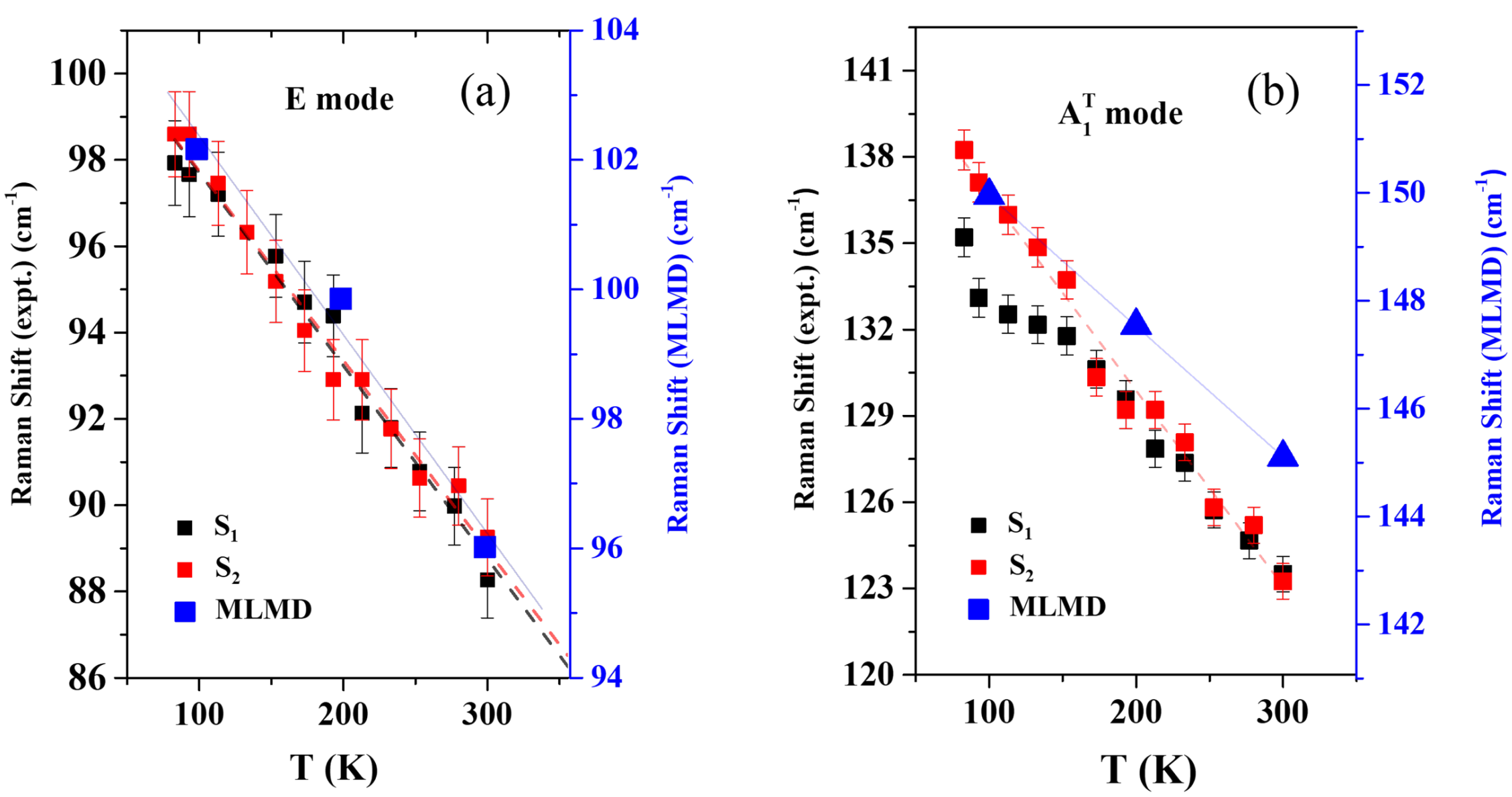

**Figure S12:** Comparison of the (a) E mode and (b) $A_1^T$ mode shift with temperature calculated using MLMD simulations with experimentally measured data

300K which depicts the E mode in Figure S12 (a) and $A_1^T$ mode in (b). The energy of the phonon modes obtained from the MLMD simulations show a similar trend which is followed the temperature dependent Raman spectra. Additionally, the rate of change in Raman shift ($\chi$ in $cm^{-1}K^{-1}$) obtained both experimentally and from simulations for the E and the $A_1^T$ mode are listed in Table S3.

Table S4: $\chi$ values for E mode and $A_1^T$ mode obtained from experimental data and MLMD simulation

| Mode | E ($cm^{-1}K^{-1}$) | $A_1^T$($cm^{-1}K^{-1}$) |
|---|---|---|
| $S_1$ | -0.04482 | Not linear |
| $S_2$ | -0.04329 | -0.0679 |
| MLMD | -0.03073 | -0.0249 |

**12. The dispersion curve for the three-phonon scattering process**

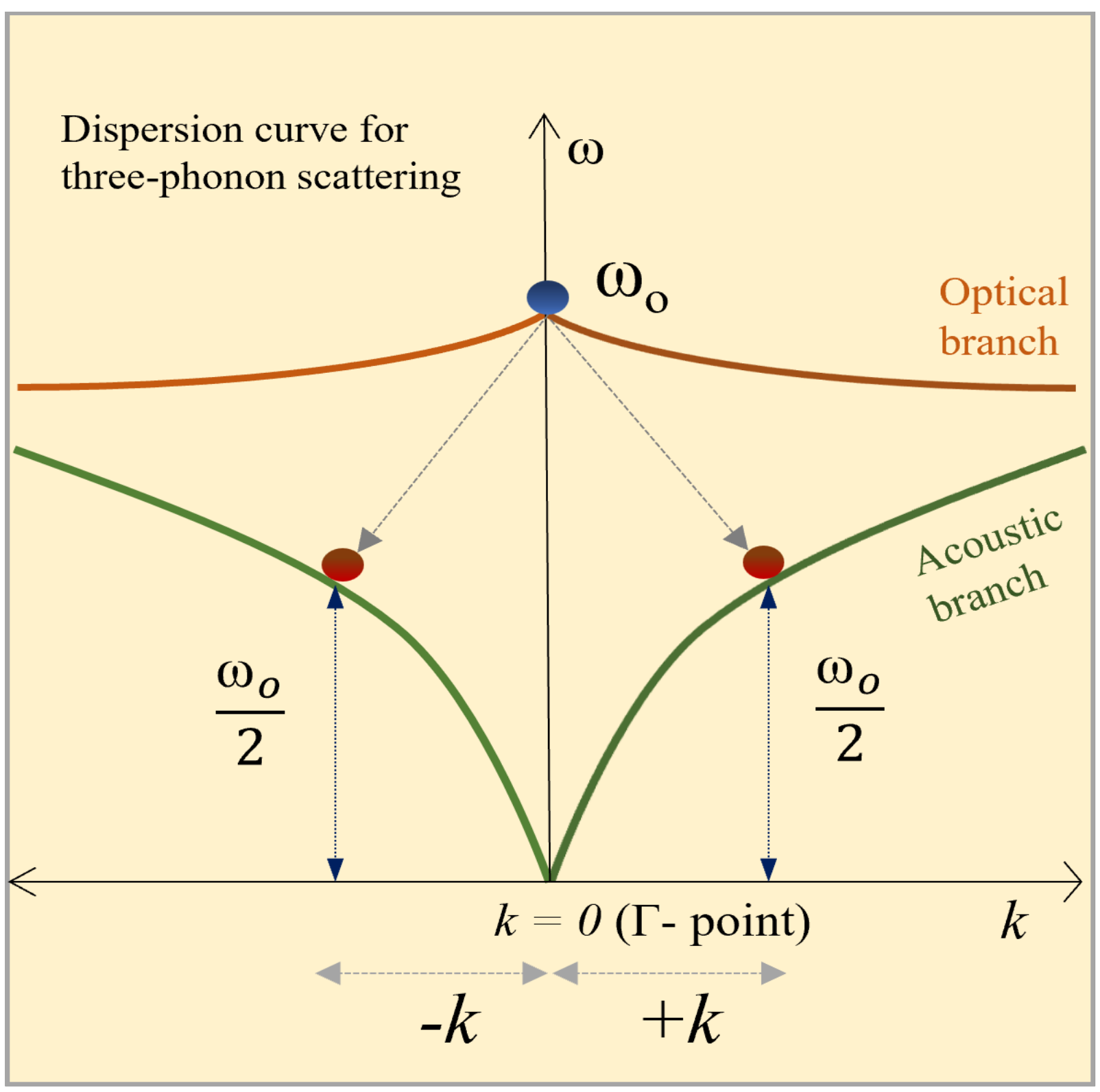

**Figure S13:** A three-phonon scattering process depicting the decay of an optical phonon of frequency $\omega_o$ at the Γ-point (k = 0) into two acoustic phonons of frequency $\omega_o/2$ having equal and opposite momentum (i.e. $\pm k$ respectively)

SUPPLEMENTARY REFERENCES